\documentclass[aps,prc,10pt,twocolumn,superscriptaddress,showpacs,longbibliography]{revtex4-2}
\usepackage{graphicx}
\usepackage{amsthm}
\usepackage{amsmath}
\usepackage{framed}
\usepackage{slashbox}
\newcommand{\cblack}{\color{black} }

\newcommand{\mev}{\, \text{MeV}}
\newcommand{\fm}{\, \text{fm}}
\usepackage[belowskip=-12pt,aboveskip=0pt,justification=justified,singlelinecheck=false]{caption}
\usepackage{caption}
\usepackage{graphics}
\usepackage{amssymb}
\usepackage{epsfig}
\usepackage{verbatim} 
\usepackage{epstopdf}
\usepackage{amsfonts}
\usepackage{transparent}
\usepackage{hhline} 

\usepackage{multirow}
\usepackage{soul}
\usepackage{subcaption}
\usepackage{cancel}
\usepackage{amsmath}
\usepackage{wrapfig}
\usepackage{float}
\usepackage{cleveref}
\usepackage{enumerate}
\usepackage{array}
\usepackage{colortbl}
{\endMakeFramed}

\newcommand{\pilesseft}{\mbox{$\pi\text{\hspace{-5.5pt}/}$}EFT$\,$}
\newcommand{\ket}[1]{\left| #1 \right>} 
\newcommand{\bra}[1]{\left< #1 \right|} 
\graphicspath{{./images_fin/}}
\begin{document}
 \pagenumbering{roman} \normalsize
 \pagenumbering{arabic}
 \cblack\title{First-principles modelling of the magnetic structure of the lightest nuclear systems using effective field theory without pions}
 \cblack
 \author{Hilla\ De-Leon}
 \email[E-mail:~]{hilla.deleon@mail.huji.ac.il}
 \affiliation{Racah Institute of Physics, The Hebrew University of Jerusalem, 
 9190401 Jerusalem, Israel}
 \affiliation{INFN-TIFPA Trento Institute of Fundamental Physics and Applications, Via Sommarive, 14, 38123 Povo TN, Italy}
 \affiliation{
 European Centre for Theoretical Studies in Nuclear Physics and Related Areas (ECT*),
 Strada delle Tabarelle 286, I-38123 Villazzano (TN), Italy}
 \author{Doron\ Gazit}
 \email[E-mail:~]{doron.gazit@mail.huji.ac.il}
 \affiliation{Racah Institute of Physics, The Hebrew University of Jerusalem, 
 9190401 Jerusalem, Israel}
 \begin{abstract}
 The strong interaction, i.e., quantum chromodynamics at the low energy nuclear regime, is notoriously known to be challenging for predictive modeling. Here, we use the simplest possible nuclear effective field theory (EFT), and show that in the case of the magnetic structure of nuclear systems with $A=2$ and $A=3$ nucleons, it is highly precise as well as predictive. The theoretical framework is the pionless EFT (\pilesseft), of point nucleons with contact interactions, expanded consistently up to next-to-leading order (NLO) in perturbation theory, i.e., including only eleven low-energy parameters, and augmented by a novel Bayesian analysis of theoretical uncertainties. The theory accurately predicts the shell structure reflected in the values of the magnetic moments and reactions of these nuclei within $\approx 1\%$ calculated theoretical uncertainty. 
 We show that this perfect prediction originates in implicit a-posteriori properties of the calculation, particularly an unexpectedly small expansion parameter, as well as a vanishing contribution from the two-body isoscalar current. 
 \end{abstract}
 \maketitle
 
 \section{Introduction}
 Low-energy magnetic transitions and magnetic moments of nuclei are sensitive to the nuclear structure and, thus, are a window into the properties of the nuclear interaction. In addition, the electromagnetic interaction shares several similarities with the weak interaction, thus providing insight into weak reactions with nuclei, which are of importance in the theoretical modeling of stellar evolution and supernovae, and are sometimes out of reach experimentally. 
Due to the non-perturbative nature of quantum chromodynamics (QCD) in the nuclear regime, it is difficult to calculate these magnetic observables directly from the more fundamental theory.
 
 Lattice formulations of QCD have only recently been able to calculate such properties, though the calculations are still restricted to unphysical quark masses~\cite{Beane:2014ora, Beane:2013br}. The latter reproduces the empirical fact that in nature and for the unphysical quark masses considered by lattice QCD calculations, the nuclear magnetic moments of nuclei of masses $A=2,\,3$ follow a perturbative trend, a shell model trend, in the sense that the Deuteron ($d$ or $^2$H) magnetic moment, $\langle\hat{\mu}_{d}\rangle$, is found to be approximately the sum of the proton and neutron magnetic moments, $\langle\hat{\mu}_{d}\rangle\approx \mu_p+\mu_n$, and the Triton ($^3$H) and Helium-3 magnetic moments are approximately the magnetic moments of the proton and neutron, respectively, consistent with a shell model behavior of a valence nucleon weakly bound to a core. 
 
 This seemingly perturbative and linear behavior is in sharp contrast with the non-perturbative quantum character of the structure of these nuclei. In the $A=2$ nuclear system, the strong interaction between nucleons leads to large nucleon-nucleon scattering lengths, or equivalently, as Bethe showed using the effective range expansion (ERE), \cite{Bethe:1949yr}, vanishing binding energy for the Deuteron, compared to the natural scale of QCD excitations. Moreover, in $A=3$ nuclei, the large nucleon-nucleon scattering length leads to a realization of the Efimov effect, making the introduction of a three-body force essential to stabilize these nuclei. 
 
 ERE can be formally phrased by using effective field theory (EFT)~\cite{3bosons, Triton}. The EFT expansion relies on scale separation in the momentum-energy regime, and thus it is natural for these nuclei since the binding momenta, $
 Q$, set by the scattering lengths, is well separated from the pion scale and related to the effective range. This is the 
 ``pionless'' version of nuclear EFT (\pilesseft), applicable with the breakdown scale set by the pion mass $\Lambda_{\rm b}\sim m_\pi \approx 140$ MeV~\cite{KSW1998_a,KSW1998_b}. The momentum scale suggests that the viable degrees of freedom eminent in determining the structure are nucleons, which interact through contact interactions. 
 \pilesseft assumes all particles, but the nucleons, are ``integrated out", and their properties dictate the size of the effective Lagrangian coefficients, called low-energy constants (LECs). The value of the LECs can be determined from experimental data~\cite{Griesshammer_pionless, few_platter} or using Lattice QCD simulations~\cite{PhysRevLett.115.132001}.
 
 Indeed, in the field theoretical formalization of \pilesseft, wave functions result from a loop integration over all the possible momenta with a cutoff $\Lambda$~\cite{faddeev, Kong1, 3bosons, Triton}. Analytical and numerical solutions of the $A=3$ integral equations show a strong dependence on the cutoff, resulting in the addition of a 3-body force counter-term, $H(\Lambda)$, already at leading order (LO), as expected from the Efimov effect~\cite{3bosons, Triton}, which ensures RG invariance. A study of realistic nuclei using \pilesseft has shown that an additional counter-term is needed to ensure RG invariance at next-to-leading order (NLO), when the Coulombic repulsion between protons is considered \cblack{\cite{konig2, K_nig_2017}}. \cblack

 In this work, we focus on the question of whether the magnetic structure of $A=2$ and $A=3$ systems can be described successfully with such a simple theory, consistent with the differences between $A=2$ and $A=3$ systems, or should many-body counter term be introduced to assure RG invariance similar to those in the strong Hamiltonian. We make use of our recently developed RG invariant method to calculate $A=3$ matrix elements within \pilesseft~\cite{Big_paper} to study \textbf{simultaneously} $A=2$ and $A=3$ $M_1$ properties beyond LO. We introduce a novel method for assessing the theoretical uncertainty due to this EFT truncation, utilizing the perturbative property of this approach. Moreover, we study different ways to implement the perturbative expansion of \pilesseft to NLO, roughly analogous to choosing different expansion points inside the convergence radius of a Taylor expansion. One of these, the Z-parameterization, shows an improved convergence pattern over the other. Lastly, we show that the statistical analysis of the $A<4$ $M_1$ observables leads to unexpected results, which are connected, as numerically argued, with the nature of the renormalization group flow of QCD to low-energies through EFTs with a larger breakdown scale, which are relevant for the study of heavier nuclei in nuclear physics.

 To this end, we examine the four well-measured low-energy magnetic ``$M_1$'' reactions, in the the $A=2,\,3$ mass nuclear systems, {\it i.e.}, the magnetic moments of the bound nuclei $\langle\hat{\mu}_{d}\rangle$, $\langle\hat{\mu}_{^3\text {H}}\rangle$ and $\langle\hat{\mu}_{^3\text{He}}\rangle$~\cite{3He_3H_data,mu_d_data}, and the cross-section $\sigma_{np}$ for the radiative capture $n+p \rightarrow d+\gamma$ for thermal neutrons~\cite{np_data}. Previous studies have used either a phenomenological approach to nuclear physics (see a recent review in Ref.~\cite{Bacca:2014tla}) or chiral EFT \cite{PhysRevC.99.034005}. These, however, do not allow to study the reactions in a consistent perturbative approach, as well as from RG perspective. Within the \pilesseft framework, a large body of work has been done on the $A=2$ aforementioned observables, with particular emphasis on $\sigma_{np}$, due to its relevance to big-bang nucleosynthesis in energy regimes characterized by large experimental uncertainties~\cite{Chen:1999bg,Rupak:1999rk,KSW_c,Chen_N_N,ando_deturon}. Recently, exploratory studies of the $A=3$ magnetic $M_1$ calculations, either to small cutoffs in a configuration space Schr\"{o}dinger equation representation of \pilesseft~\cite{Kirscher:2017fqc}, or without including the Coulomb interactions for $^3$He~\cite{Vanasse:2017kgh}. As a result of the approximations, these studies could not approach the aforementioned questions.

 \section{Setting up \pilesseft to next-to-leading order}\label{two_body}
 
 At LO, the two-body \pilesseft Lagrangian is minimally built with SU(2) nucleon fields to reproduce the bound spin-triplet channel ($t$, Deuteron) binding momentum, $\gamma_t\sim\mathcal{O}(Q)$ and the unbound spin-singlet channel, $s$, scattering length, $a_s \sim \mathcal{O}({1}/{Q})$). These values are unnatural compared to the QCD excitations (specifically, compared to the pion mass) and are the signature of the ``strength'' of the strong interaction, i.e., of the unnaturally large cross-section compared to the nuclear matter-radius. Details of the interaction enter at NLO, through its effective range in the triplet and singlet channels, $\rho_t$ and $\rho_s$, respectively, which scale as $\mathcal{O}\left({1}/{\Lambda_{\rm b}}\right)$. Clearly, the nuclear system at low energies is characterized by the properties of two-body clusters; thus, it is convenient to use a Hubbard-Stratonovich (H-S) transformation to equivalently reformulate \pilesseft with dynamical dinucleon fields alongside with the nucleon field~\cite{rearrange, Bedaque:1999vb}. 
The fields $t$ and $s$ have quantum numbers of two coupled nucleons in an S-wave spin-triplet and -singlet state, respectively. This simplifies the calculation of a three-body amplitude by turning it into an effective two-body scattering problem of a dinucleon and a nucleon (see, for example, Refs.~\cite{rearrange, Bedaque:1999vb}).
 
 This explanation entails that the effective ranges vanish at LO, and receive a finite value at NLO. For the unbound spin-singlet state the natural choice is called effective-range (ER) parameterization in which $\rho_s=\underbrace{0}_{\text{LO}}+\underbrace{\rho_s^{\text{exp}}}_{\text{NLO}}$, where $\rho_s^{\text{exp}}$ is the experimental value. However, for the bound triplet channel, there is another useful choice for its effective range value at NLO. Since it is bound, the long-range properties of the Deuteron wave function can be set by a quantity $Z_d$, defined through the Deuteron asymptotic S-state normalization, ${A}_S$, such that ${A}_S\equiv\sqrt{2\gamma_t Z_d}$. $A_s$ roughly dictates the long range normalization of the deuteron wave function, 
 $Z_d = \frac{1}{1-\gamma_t \rho_t} \approx 1 + \gamma_t \rho_t$. 
 
 In the ER parameterization, $\rho_t$ is fixed to its physical value already at NLO, which is reflected in a $\sim17\%$ deviation of $Z_d$ from its physical value at NLO. The alternative arrangement of the \pilesseft is called Z-paramereization, and it fixes $Z_d$ to its experimental value at NLO, i.e.,  $Z_d^{\text{NLO}}=Z_d^{\text{exp}}\approx 1.690(3)$, while $\rho_t$ is subsequently derived to be, $\rho_t^{Z}\approx\underbrace{0}_{\text{LO}}+\underbrace{\frac{Z_d^{\text{exp}}-1}{\gamma_t}}_{\text{NLO}}= 2.976 (5)\fm$, deviating significantly by about 40\% (!) from its physical value (see Tab.~\ref{table_exp_data}). 
 ~\cite{Phillips_N_N,Griesshammer_3body,Kong2,Phillips:1999am,Vanasse,Vanasse:2015fph}.
 It was shown that this choice recovers elastic scattering data better and has a faster convergence in the EFT expansion ~\cite{Phillips:1999am, Rho:1999bv}.
 
 \begin{table}[h]
 \begin{center}
  \begin{tabular}{cc||cc}
  \multicolumn{2}{c||}{LO}&\multicolumn{2}{c}{NLO}\\
  \hline
  Parameter& Value 
  & Parameter& Value \\
  \hline
  {$\gamma_t$}& { 45.701 MeV}& $\rho_t$ &1.765 fm (ER)\cite{33}\\
  $a_s$& -23.714 fm \cite{Preston_1975}& 
  $\rho_s$& 2.73 fm \cite{deSwart:1995ui}\\
  $a_p$& -7.8063 fm \cite{34} &
  $\rho_C$& 2.794 fm \cite{34}\\
  $\kappa_0$ & 0.4399.. NM&&
  \\
   $\kappa_1$ & 2.3529.. NM&&
  
  \end{tabular}
  \caption{ \footnotesize{Experimental parameters used in the calculation. Note that the calculation additionally includes a three-body force parameter at LO, an isospin dependent three-body force at NLO, and two magnetic two-body current parameters at NLO.}}
  \label{table_exp_data}
 \end{center}
 \end{table}

 In the following, we use both NLO parameterizations, check that they are RG invariant, and compare them in the context of $A<4$ $M_1$ observables.
 
 The three-nucleon scattering amplitude results from the full solution of the coupled channel Faddeev integral equations. The different channels for $^3$H are the spin-triplet - $t$ (representing an ``off-shell'' Deuteron, $d$, dibaryon), and the spin-singlet - $s$ ($nn,np$). $^3$He is characterized by spin-triplets - $t$, spin-singlets - $s$ ($np$), and $pp$~\cite{faddeev}, the latter required by the Coulomb force between protons, which is fully considered by this work. The nuclear amplitudes we use here are taken explicitly from Ref.~\cite{Big_paper}, where they are benchmarked numerically and validated using the binding energy difference between $^3$H and $^3$He.

 \section{$M_1$ observables in the $A<4$ systems}\label{M1_observables}
 
 $M_1$ observables at vanishing momentum transfer are related to the electromagnetic nuclear current density $\mathcal{\hat{J}} (\vec{q})$. 
 Explicitly, the magnetic moment of a bound state is just the expectation value of the operator:
 \begin{equation} \label{Eq:mu_def}
 \hat{\mu}=-\frac{i}{2} \vec{\nabla}_q \times \mathcal{\hat{J}} (\vec{q})\big{|}_{q=0},
 \end{equation}
 while $\sigma_{np}$ is proportional to the transition matrix element of the same operator between the neutron and proton, $S=0$ state, to the Deuteron, $S=1$ state~\cite{rearrange,Kaplan:1998xi,ando_deturon}.

 A magnetic photon interaction with a nucleus can be modeled effectively as an interaction with ever-growing clusters of nucleons. In \pilesseft, LO includes a single nucleon interaction with a photon, while the interaction of a magnetic photon with two-body clusters appears for the first time at NLO~\cite{rearrange, KSW_c,ando_deturon}.

 The one-body electromagnetic Lagrangian is given by (see, for example, Ref.~\cite{Chen_N_N}): 
 \begin{equation}\label{eq_l_magntic_1}
 \mathcal{L}^{\text{1-B}}_{\text{magnetic}}=\frac{e}{2M}N^\dagger\left (\kappa_0+\kappa_1\tau_3\right)\vec{\sigma}\cdot \vec{B} N~,
 \end{equation}
 where $\vec{B}$ is the magnetic field, $e$ is the electron charge, while $\kappa_0$ and $\kappa_1$ are the isoscalar and isovector magnetic moments of the nucleon. 
 
 The NLO interaction of a magnetic photon with a two-body nuclear field is given by the two-body electromagnetic Lagrangian in the form of two four-nucleon-one-magnetic-photon operators: 
 \begin{equation}\label{l_magntic_2}
 \begin{split}
 \mathcal{L}^{\text{2-B}}_{\text{magnetic}}=&e\left[ L'_1\left (N^TP_s^AN\right)^\dagger\left (N^TP_t^iN\right)B_i\right.\\
 &-\left.L'_2\left (N^TP_t^iN\right)^\dagger\left (N^TP_t^jN\right)B_k+h.c\right]~.
 \end{split}
 \end{equation}
 where $P_{t,s}^{i,A}$ are projection operators on the $t$ and $s$ channels.

 Applying the H-S transformation to \cref{l_magntic_2} leads to the interaction in terms of the dibaryon fields (see~\cite{ando_deturon,Big_paper}): 
 \begin{equation}\label{eq_magnetic_dibaryon}
 \begin{split}
 &\mathcal{L}^{\text{2-B}}_{\text{magnetic}}=\\&\frac{e}{2M}\left[\kappa_1L_1 (t^\dagger s+s^\dagger t)\cdot \vec{B}-i\epsilon^{ijk}\kappa_0L_2 ((t^i)^\dagger t^j)\cdot {B}_k\right]~.
 \end{split}
 \end{equation}
 The LECs $L_1$ and $L_2$ can be separated to a LO part, which is related to a consistent ERE, and NLO pure two-body contributions~\cite{ando_deturon,ando_magntic_BBN},
 \begin{eqnarray}
 L_1 (\mu)&=&\underbrace{-\frac{\rho_t+\rho_s}{\sqrt{\rho_t\rho_s}}}_{\text{LO}}+\frac{4}{\gamma_t\sqrt{\rho_t\rho_s}}\underbrace{l'_{1}(\mu)}_{\text{NLO}}~ \\
 L_2(\mu)&=&-\underbrace{2}_{\text{LO}}+\frac{2}{\gamma_t\rho_t}\underbrace{l'_2(\mu)}_{\text{NLO}}~,
 \end{eqnarray}
 with $l'_{1}, l'_{2}$ being renormalization scale independent for $\mu\rightarrow \infty$. In this work, contrary to past studies on the electromagnetic properties of light nuclei, which arbitrarily took $\mu=m_\pi$ (see, for example, Ref.\cite{Chen:1999tn}), we check the full renormalizability to essentially infinite cutoffs, and use the value of the parameters at \cblack $\mu=\Lambda$ (the ultraviolet cutoff) $\rightarrow \infty$ for both $A=2$ and $A=3$ \footnote{The purpose of this work is to examine the consistency of pionless EFT for the transition from $A=2$ to $A=3$ and vice versa. Hence, to eliminate any artificial effects, we choose to use the same value $\left(\mu=
 \Lambda\rightarrow \infty\right)$ for both the two-body dimensional regularization and the three-body cutoff regularization, since we expect that $\lim_{\mu=
 \Lambda\rightarrow\infty}\frac{d M_1(\mu)}{d\mu}=0$ for both A=2 and A=3 observables}\cblack. We note that since the photon field $\mathcal {\vec{A}}$ fulfills $\vec{B}=\vec{\nabla}\times {\mathcal{\vec{A}}} (\vec{x})$, the scattering operator $\hat{\mu}$ is given by the prefactor of $\vec{B}$ in \cref{eq_l_magntic_1,eq_magnetic_dibaryon}. Feynman rules are extracted trivially using this fact.
 
 Given the above, the $A<4$ $M_1$ observables are calculated consistently up to NLO in the following way~\cite{Big_paper}:
 \begin{multline}\label{eq_NLO_parts}
 \langle \hat{\mu}\rangle=\langle \hat{\mu}\rangle^{\text{1-B}}_{\text{LO}}\times\\\left(\underbrace{1}_{
 \text{LO}}+\underbrace{\delta\langle\hat{\mu}\rangle^{\text{1-B}}_{\text{ERE}}+\delta\langle\hat{\mu}\rangle^{\text{2-B}}_{\text{ERE}}}_{\substack{ \text{LO magntic opert.}\\\text{NLO storng inter.}}}+\underbrace{\delta\langle\hat{\mu}\rangle^{\text{2-B}}}_{\substack{ \text{NLO magntic opert.}\\\text{LO storng inter.}}}\right)~.
 \end{multline}
 \subsection{The $A<4$ $M_1$ matrix elements}
 
 \subsubsection{Two-nucleon electromagnetic matrix elements}
\begin{figure}[h]
 \vspace{-0 cm}
 \centering
 \includegraphics[width=0.99\linewidth]{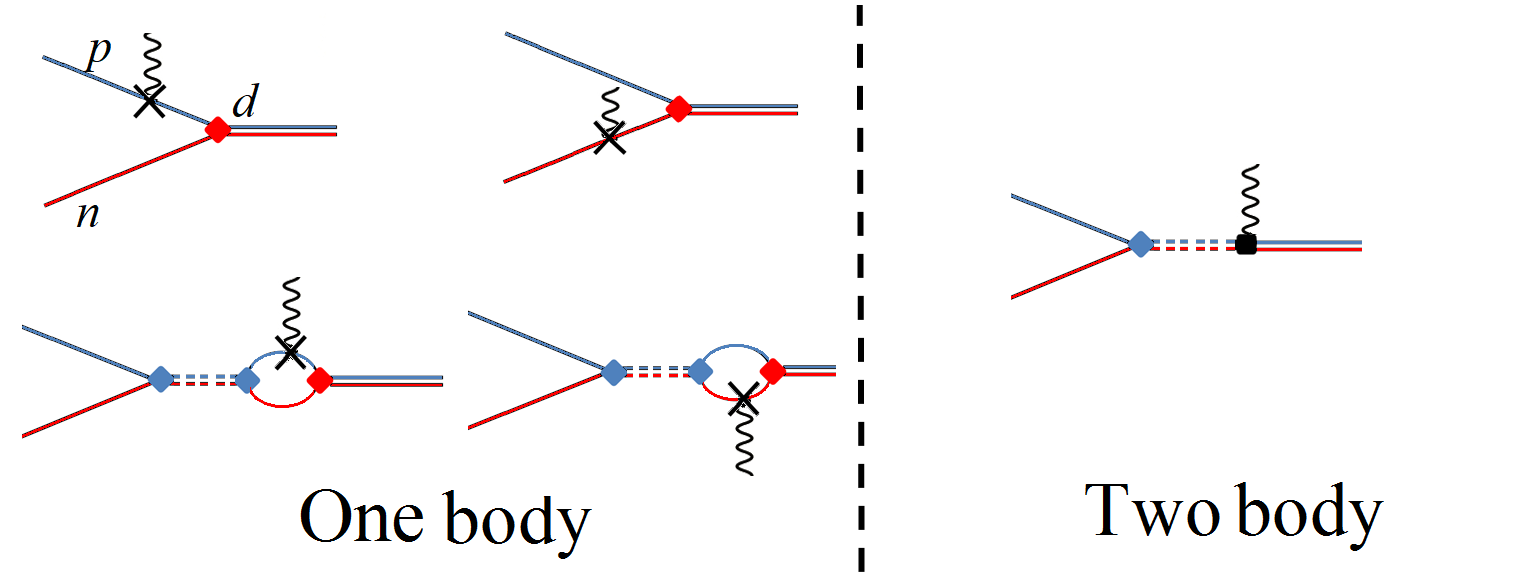}
 \captionsetup{justification=raggedright, singlelinecheck=false}
 \caption{\footnotesize{Diagrammatic representation of the calculation of $\hat{\mu}$ between two-body states, up to NLO. $\hat{\mu}$ insertion is represented by the photon vertex. The double lines are the NLO propagator of the two dibaryon fields. $D_t$ (solid) and $D_s$ (dashed). The red (blue) line represents the neutron (proton) propagator. A spin-singlet (triplet) dibaryon-nucleon-nucleon vertex represents a strong interaction between pairs of nucleons.}}
 \label{fig_np_capture}
 
 \end{figure}

 The matrix element of $\hat{\mu}$ (\cref{Eq:mu_def}) between two-nucleon states is represented diagrammatically in Fig.~\ref{fig_np_capture}. This matrix element is related to the calculation of $\sigma_{np}$ ($\langle\hat{\mu}_{d}\rangle$) if the initial state is in a relative $^1\text{S}_0$ ($^3\text{S}_1$) state. The field $^3S_1$ state represents the deuteron.
 
 From Fig.~\ref{fig_np_capture}, one concludes that up to NLO, $\langle\hat{\mu}_{d}\rangle$ is given by: 
 \begin{multline} \label{eq_mu_fineal}
 \langle\hat{\mu}_{d}\rangle =\kappa_0\left\{2Z_d^{\text{NLO}}+Z_d^{\text{LO}}\left[\gamma_t\rho_t L_2 (\mu)\right]\right\}=\\
 =2\kappa_0\left[1+\underbrace{0}_{\text{NLO strong inter.}}+\underbrace{l'_2 (\mu)}_{\substack{ \text{NLO magnetic opert.}\\\text{LO strong inter.}\\}}
 \right],
 \end{multline}
 
 The cross-section of $n+p\rightarrow d+\gamma$ is related to the matrix element $Y$ by: 
 \begin{equation}\label{eq_sigma}
 \sigma_{np} =2\alpha\pi \dfrac{\left (\gamma_t^2+q^2/4\right)^3a_s^2}{ M^4q\gamma_t}Y_{np}^2 \equiv 2\alpha\pi \dfrac{\gamma_t^5a_s^2}{ M^4q} (2\kappa_1)^2 (Y'_{np})^2, 
 \end{equation}
 where $Y_{np}$ is the sum over all the diagrams of Fig.~\ref{fig_np_capture} and $q=0.0069 \mev/c$ is the momentum transfer for thermal neutrons~\cite{Vanasse:2017kgh,PhysRevLett.115.132001}, \footnote{In this work, similar to Refs.~\cite{Vanasse:2017kgh,PhysRevLett.115.132001}, we use $q=2Mv_{\text{lab}}=2\cdot M 2200 \text{M/s}=0.0069\mev$. This value is higher than the value used in Refs.~\cite{Chen:1999tn,ando_deturon},$q=0.068\mev.$ We found that $l_1' (q=0.0069\mev)$ is higher than $l_2' (q=0.0068\mev)$ by 20\%(10\%) for the Z-(ER-) parameterization}.

 The normalized matrix element, $Y'_{np}$, up to NLO is also obtained by Fig.~\ref{fig_np_capture} to yield:
 \begin{multline} \label{eq_Y_fineal}
 Y'_{np} =\left (1-\frac{1}{\gamma_ta_s}\right)\times
 \Bigg[1+\Bigg.\\
 \Bigg.\underbrace{\sqrt{Z_d^{\text{NLO}}}-1-\frac{\gamma_ta_s}{\gamma_ta_s-1}\frac{\gamma_t\left(\rho_t+\rho_s\right)}{4}}_{\substack{\text{NLO storng inter. corrections}\\\text{LO magnetic op.}}}+\underbrace{\frac{\gamma_ta_s}{\gamma_ta_s-1}l'_1 (\mu)}_{\substack{\text{\parbox{2cm}{\centering NLO magnetic \\[-4pt] op. corrections}}\\{\,}\\\text{\parbox{2cm}{\centering LO strong inter.\\[-4pt]}}} 
 }\Bigg],
 \end{multline}

 The above expressions up to higher-order corrections can be found in the literature (see, e.g.,~\cite{Chen:1999tn, ando_deturon}). 
 \subsubsection{Three-nucleon electromagnetic matrix elements}
 
 In Ref.~\cite{Big_paper}, we presented a general perturbative diagrammatic approach for calculating one- and two-body matrix elements between initial and final three-nucleon bound-states up to NLO. In this work, we use this method to calculate the $M_1$ observables in the $A=3$ system (see Appendix~A). The $A=3$ $M_1$ observables, can be separated to different contributions, similarly to Eq.~\ref{eq_NLO_parts},
 
 though calculated numerically (using the experimental input parameters shown in Tab.~\ref{table_exp_data}), see Ref.~\cite{Big_paper}.
 
 \section{Results and analysis}\label{results}
 
 The $A<4$ $M_1$ observables depend upon the physical (RG invariant) values of the LECs, {\it i.e.,} ${l'_{1,2}}^\infty\equiv l'_{1,2} (\mu=\Lambda\rightarrow \infty)$. In past works, the experimental values of the $A=2$ observables ($\sigma_{np}$ and $\langle\hat{\mu}_d\rangle$) were used to fix these LECs~\cite{ando_deturon, ando_magntic_BBN}. Here, we calculate consistently the $A<4$ $M_1$ observables 
 which depend on the same LECs, so we can extract these LECs from two observables and then use them to predict the remaining 
 two observables. Therefore, we have six independent ways for calibrating the LECs. These calibrations are used to check whether this formalism can be consistently used to describe simultaneously $A=2$ and $A=3$ 
 $M_1$ observables using \pilesseft. 
 
 Tab.~II~(a) summarizes our predictions for ${l'_{1,2}}^\infty$ and $M_1$ observables up to NLO {in both $Z$- and ER- parameterization}. For each row, the $\star'$ denotes the $M_1$ observables used to calibrate ${l'_{1,2}}^\infty$. For example, the first row of Tab.~II~(a) shows the LECs fixed from $A=3$ observables and our prediction of $A=2$ magnetic observables, while the second row of Tab.~II~(a) shows the LECs fixed from $A=2$ observables and the prediction of $A=3$ magnetic observables. 
 Note that for each $M_1$ observable, we have three predictions.
 
\begin{widetext}

 \begin{figure}[h!]
  \begin{center}
  \includegraphics[width=1\linewidth]{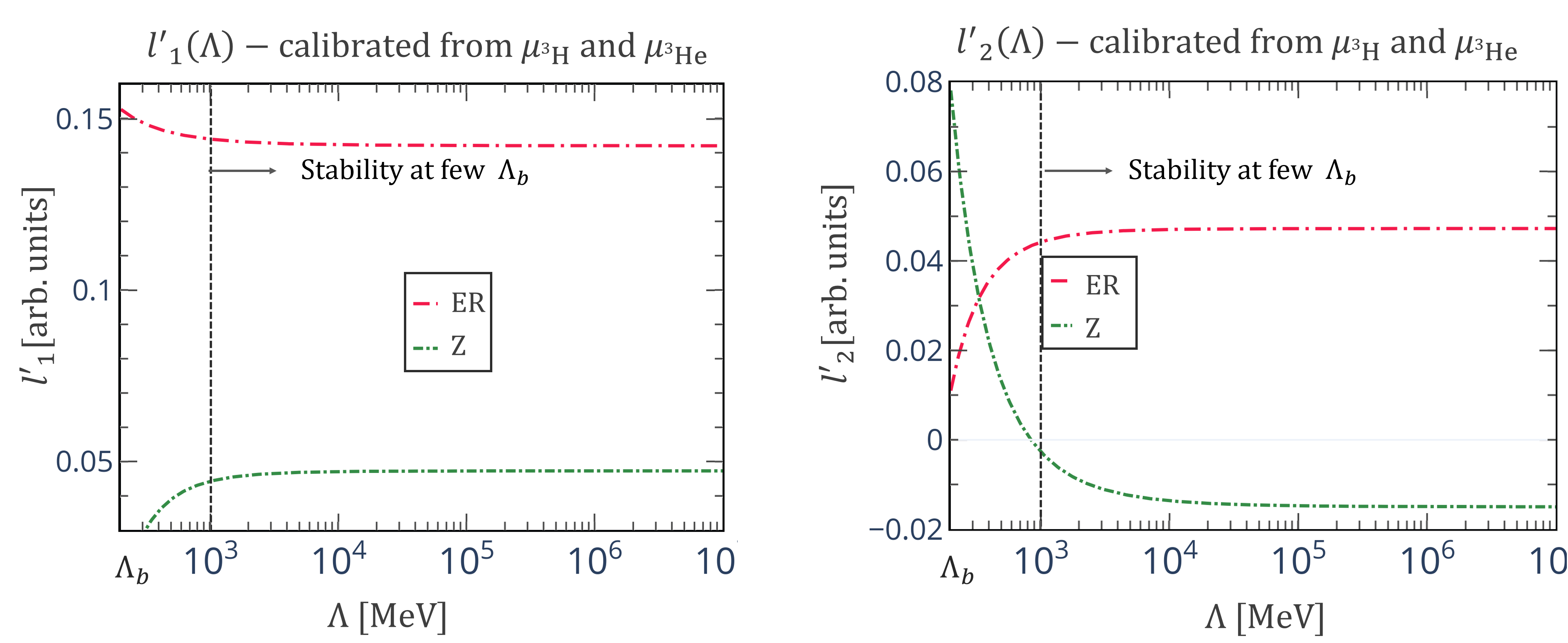}\\
  \caption{\footnotesize{
   Numerical results for the LECs $l'_1(\Lambda)$ (left panel) and $l'_2(\Lambda)$ (right panel), calibrated from the $M_1=3$ observables as a function of the cutoff $\Lambda$. The long (short) dotted-dashed lines are the numerical results 
   for the ER-(Z-) parameterization.}}\label{fig_L1_L2}
  \end{center}
 \end{figure} 
 \end{widetext}

 The numerical results of ${l'_1}(\Lambda)$ and ${l'_2}(\Lambda)$ are shown in Fig.~\ref{fig_L1_L2}, and show that they become RG invariant at $\mu \sim$ few $\Lambda_{\rm b}$, if they are fixed in the $A=3$ systems. 
 
  \begin{widetext}
 \begin{center}
  \begin{table}[H] 
  
  \begin{tabular}{lr}
   
   \begin{tabular}{c|| c| c|c|c|c|c}
   & ${l'_{1}}^{\infty}/ 10^{-2}$&${l'_{2}}^{\infty}/ 10^{-2}$&$\langle\hat{\mu}_{^3\text{H}}\rangle$[NM]& $|\langle\hat{\mu}_{^3\text{He}}\rangle|$[NM]&
   $\langle\hat{\mu}_{d}\rangle$[NM]&$Y'_{np}$\\ 
   \hhline{=|=|=|=|=|=|=}
   LO&0 (0)& 0 (0)& $2.76$ $(2.78)$ & $1.84$ $(1.84)$ & 0.88 (0.88) &1.18 (1.18) \\
   \hline
   \multirow{6}{*}{\rotatebox[origin=l]{90}{\parbox[l]{.9cm}{NLO}}}&4.72 (14.2)& -1.6 (4.1)& $\star$ & $\star$ & 0.87 (0.92) &1.253 (1.31) \\
   &4.66 (9.0) &-2.6 (-2.6) & 2.978 (2.76) & 2.145 (1.89) & $\star$ &$\star$ \\
   &4.66 (9.0) & -2.4 (29) & $\star$ & 2.144 (1.66) & 0.86 (1.17) &$\star$ \\
   &4.66 (9.0) & -0.13 (-31) & 2.996 (2.59) & $\star$ & 0.88 (0.61) &$\star$ \\
   &4.92 (15.2)& -2.6 (-2.6)&$\star$ & 2.143 (2.23) & $\star$ &1.255 (1.32) \\
   &4.60 (13.4)& -2.6 (-2.6)& 2.967 (2.91) & $\star$ & $\star$ &1.253 (1.30) \\
   \hline 
   \multicolumn{1}{c||}{Mean} & 4.73 (13.0) & -1.7 (-0.04)& 2.98 (2.75) &2.144 (1.93)& 0.87 (0.89)& 1.253 (1.31) \\
   \hline 
   \multicolumn{1}{c||}{std}&0.2 (2.8) & 1.1 (25) &0.015 (0.16)& 0.001 (0.28) &0.01 (0.26)& 0.001 (0.01) \\
   \hline
   \hline
   \multicolumn{1}{c||}{Exp}&&&&&&\\
   \multicolumn{1}{c||}{data}&&& \multirow{-2}{*}{2.979~\cite{3H_3He_magnetic}}& \multirow{-2}{*}{2.128~\cite{3H_3He_magnetic}}& \multirow{-2}{*}{0.857~\cite{mu_d_data}}& \multirow{-2}{*}{1.253~\cite{np_data}}
   \end{tabular}
   &
   \begin{tabular}{c| c| c|c}
   $M_1$&$\delta\langle\hat{\mu}\rangle_{\text{total}}$&$\delta\langle\hat{\mu}\rangle_{\substack{\text{NLO}\\{\text{strong}}\\{\text{inter.}}}}$&$\delta\langle\hat{\mu}\rangle^{ \text{2-B}}_{\substack{\text{NLO}\\{\text{magnetic}}\\{\text{opert.}}}}$\\ 
   \hhline{=|=|=|=}
   $\langle\hat{\mu}_{^3\text{H}}\rangle$&7\% (1\%)&3\% (11\%)&5\% (10\%)\\
   $\langle\hat{\mu}_{^3\text{He}}\rangle$&13\% (4\%)&3\% (25\%)&10\% (29\%)\\
   $\langle\hat{\mu}_{d}\rangle$& 1\% (1\%)&0\% (0\%)&1\% (1\%)\\
   $Y'_{np}$ &6\% (9\%)&2\% (2\%)&4\% (12\%)\\
   \multicolumn{4}{c}{}\\
   \multicolumn{4}{c}{}\\
   \multicolumn{4}{c}{}\\
   \multicolumn{4}{c}{}\\
   \multicolumn{4}{c}{}\\
   \multicolumn{4}{c}{}\\
   \end{tabular}
  \end{tabular}
  \caption{ \footnotesize{\begin{minipage}{0.55 \linewidth}
    (a) Numerical results for ${l'_{1}}^{\infty},{l'_{2}}^{\infty}$ and $A=2,3$ $M_1$ observables. The nominal value is calculated using $Z$-parameterization, while the number in brackets is calculated using the ER-parameterization. ``Mean" denotes the mean value of the $M_1$ observable based on its three (independent) predictions, while ``std" denotes the standard deviation of these independent predictions.
   \end{minipage}\qquad
   \begin{minipage}{0.33 \linewidth}
    (b) The order-by-order contributions to the $M_1$ matrix elements, based on their mean values given in Tab.~II~(a). The nominal value is calculated using $Z$-parameterization, while the number in brackets is calculated using ER-parameterization. The three nuclear magnetic moments are given in nuclear magnetons [NM] 
  \end{minipage}}}
  \end{table}
  \label{Tab_NLO_parts}
 \end{center}
 \end{widetext}
 In addition, in Tab.~II~(b) we present the three contributions to the different $M_1$ observables for both the ER- and $Z$-parameterizations.

 Our results verify that the \pilesseft calculations presented in this paper are purely perturbative \cblack{in the ERE expantion}\footnote{Here, since we are focusing on the ERE order-by-order expansion of the matrix element, we took the full Coulomb descriptions of $^3$He. It should be noted that in another work that focused on weak interaction \cite{de2019tritium}, we examined the effects of Coulomb interaction on the weak matrix element and showed thing at it is significantly lower in comparison to that originates from the ERE.}, \cblack {\it i.e.,} consistently organizing the expansion in a perturbative manner, and built theoretically without including any higher-order terms. Moreover, an order-by-order renormalization was obtained, as shown numerically, by the cutoff invariance (see Fig.~\ref{fig_L1_L2}) with a small expansion parameter of about $0.05-0.2$. In chiral effective field theory ($\chi$EFT), as well as in \pilesseft configuration space schemes~\cite{2016PhLB..755..253K, Kirscher:2017fqc}, a cutoff variation is frequently used to obtain an uncertainty estimate \cblack{ which is a perfectly valid procedure in a renormalizable
theory}. 
Here we show that the main advantage of using the current formalism of
 \pilesseft is the cutoff invariance, which even for $A=3$ calculations is obtained at a natural scale \footnote{Cutoff dependence for low cutoff is expected since the cutoff functions inflicted on matrix element and wave functions cannot be made fully consistent.}. This cutoff independence not only removes questions regarding residual cutoff dependencies that might contribute to the total uncertainty~\cite{Kirscher:2017fqc, 2016PhLB..755..253K}, but also allows giving physical meaning to the size of the NLO contribution.

 \subsection{ Z-parameterization is better than the ER-parameterization at NLO} 
 
 The comparison between the results of the Z- and ER-parameterizations reveals some interesting features. The ratios between the NLO and LO values are of the same order of magnitude, slightly smaller in the ER-parameterization. Na\"{i}vely, this can be interpreted as an indication of a better convergence pattern of the ER-parameterization. However, a closer look shows the contrary.
 First, in the ER-parameterization, one observes a large cancellation between the different contributions to the NLO, \textit{i.e.,} between the range corrections and the pure two-body contact contributions. Each of the NLO contributions is usually more than 10\% of the LO, while the final NLO contribution is an order of magnitude smaller. The
 Z-parameterization shows a natural convergence pattern, where all the contributions are large of the same order of magnitude.
 \cblack{Also, for the Z-parametrization, the total NLO value converges faster to the observational value. Therefore, to maintain a logical dependence on expansions orders, we assume that the fast convergence implies that up to NLO, the Z-parametrization is the preferred parametrization.} \cblack
 Second, and as a consequence of the former point, the resulting statistical standard deviations (std) between the predictions using the different LECs calibrations (see Tab.~II~(a)) for the four magnetic
 observables are an order of magnitude bigger in the ER-parameterization than the Z-parameterization.
 Third, the statistical fluctuations in the LECs' sizes, as seen in Tab.~II~(a), are much bigger in the ER-parameterization. 
 
 The large variations and fluctuations of the ER-parameterization raise questions about its relevance at NLO for predictions of electromagnetic observables. The advantage of Z-parameterization over ER-parameterization at NLO, as explicitly demonstrated here in the magnetic properties of the $A<4$ systems, is consistent with the initial motivation for introducing Z-parameterization~\cite{Phillips:1999am, Rho:1999bv, Griesshammer_3body}.
 In the next sub-sections, we examine the \pilesseft NLO's contributions and estimate its truncation error only for the Z-parameterization, which shows a more natural convergence pattern and is expected to have better predictive power.
 
 \subsection{Isoscalar two-body coupling is consistent with zero}
 
 Interestingly, we find that while ${l'_{1}}^{\infty}$ has minor dependence on the $M_1$ observables used for its calibration, \textit{i.e.,} $ \Delta{l'_{1}}^{\infty}/{l'_1}\approx3\%$, the standard deviation of ${l'_{2}}^{\infty}$ is of the same order of magnitude as ${l'_{2}}^{\infty}$, \textit{i.e.,} $ \Delta{l'_{2}}^{\infty}/{l'_{2}}^{\infty}\approx70\%$, in the Z-parameterization. The differences are even more significant in the case of ER-parameterization, where $\Delta{l'_{1}}^{\infty}_{ER}/{l'_{1}}^{\infty}_{ER}\approx21\%$, and $\Delta{l'_{2}}^{\infty}_{ER}$ is two orders of magnitude larger than ${l'_{2}}^{\infty}_{ER}$. 
 
 Moreover, the NLO contribution to the Deuteron magnetic moment is very small; in fact, it is much smaller than the NLO contributions of the other observables. The two-body
 contribution to $\hat{\mu}_d$, as seen in \cref{eq_mu_fineal}, depends only on ${l'_2}^\infty=\left(-1.7\pm1.1\right)\cdot10^{-2}$. 
 
 These two observations show that ${l'_2}^\infty$ is consistent with zero. As an interpretation, the isoscalar pure two-body contribution, whose LEC is $[{l'}_2^\infty$, might be regarded as a higher order than NLO, in contrast to the naive dimensional analysis of /pilesseft, where both the two-body isovector and isoscalar contributions are treated as NLO~.\cite{Chen:1999tn,rearrange}.
 
 To check the consistency of our conjecture, we study the ramifications of vanishing ${l'}_2^\infty$. Similarly to Tab.~II~(a), for each row in Tab. \ref{table_all_M1_l2_0}, the $'\star'$ denotes the $M_1$ observable used for ${l'_1}^\infty$ calibration.
 \begin {table}[H]
 \centering
 \begin{tabular}{c| c| c|c|c}
 & ${l'_{1}}^{\infty}/ 10^{-2}$&$\langle\hat{\mu}_{^3\text{H}}\rangle$[NM]
 & $\langle\hat{\mu}_{^3\text{He}}\rangle$[NM]
 &$Y'_{np}$\\ 
 \hhline{=|=|=|=|=}
 &4.36 & $\star$ &-2.10 &1.250\\
 &4.97 &3.00 &$\star$ &1.256\\
 &4.66 &2.99 &-2.11 &$\star$\\
 \hline 
 \multicolumn{1}{c|}{Mean} & 4.7 &2.99&-2.11 &1.253\\
 \hline 
 \multicolumn{1}{c|}{Standard deviation} &0.6 &
 0.01
 &0.01 &0.006 \\
 \hline
 \multicolumn{1}{c|}{$\%$NLO/LO} & &8\% &14\% & 6\% \\\multicolumn{1}{c|}{Exp. data}& & 2.979 & -2.128 & 1.253
 \end{tabular}
 \vspace{0.2 cm}
 \caption{ \footnotesize{Numerical results for the calibrated values of ${l'_1}^\infty$ and the resulting predictions of $M_1$ observables up to NLO for the $Z$-parameterization for the case that ${l'_{2}}^{\infty}=0$. }}
 \label{table_all_M1_l2_0}
\end{table}

Table.~\ref{table_all_M1_l2_0} shows that setting ${l'_2}^\infty$ to zero does not reduce the quality of ${l'_1}^\infty$ and $M_1$ predictions in terms of the size of the NLO contribution compared to the LO one, and the statistical accuracy of the predictions given different experimental constraints.
This implies that there is no inconsistency in assuming that the isoscalar ${l'_2}^\infty$ contribution to the $M_1$ matrix elements is suppressed compared to NLO contributions. This is a main result of this paper, and it will be discussed further in sub-section~\ref{estimation}. One is also tempted to compare the predictions to the experimental values. However, for a complete comparison, one needs to estimate the theoretical uncertainty, which is the subject of the
next sub-section.

\subsection{Estimating theoretical uncertainty}\label{estimation}
The aforementioned fact that EFT is a systematic expansion in some small parameter, $\delta$, is particularly helpful for estimating theoretical uncertainties in the calculation. A common approach is to study the residual cutoff dependence and to use it as a measure for the uncertainty (see, e.g.,~\cite{Kirscher:2017fqc,2016PhLB..755..253K} in the context of \pilesseft). The order-by-order renormalization group invariance achieved at a few times the physical breakdown scale in this work removes this source of uncertainty.

An additional approach to estimating the theoretical uncertainty is studying the truncation error in the systematic EFT expansion~\cite{PhysRevC.92.024005}. In order to do so, let us write the \pilesseft expansion for any $M_1$ observable as:
\begin{equation} \label{eq: expansion}
\langle{M_1}\rangle=\langle{M_1}\rangle_{\text {LO}} \cdot \left (1+\underbrace{c^{\text {NLO}}_{M_1}\cdot\delta}_{\mathcal{O}\left(\text{NLO}\right)}+ {\mathcal{O}} (\delta^2)\right).
\end{equation}
EFT suggests that $c^{\text {NLO}}_{M_1}$ is of natural size, and thus the truncation error is dictated by $\delta$. In \pilesseft, the na\"ive expansion parameter is estimated from $\delta\approx \frac{\gamma_t}{m_\pi} \approx \frac{1}{3}$. Using this expansion parameter, $\delta$, is usually the starting point for estimating theoretical uncertainties~\cite{Griesshammer:2015ahu}. Here, we estimate the expansion parameter, $\delta$, directly from the calculations (see Refs.~\cite{PhysRevC.99.034005,PhysRevC.92.024005} for other attempts to estimate the expansion parameter, though within $\chi$EFT). 

Powers of $\delta$ appear in ratios of perturbation expansion orders. The calculations here include a few such probes of $\delta$, including the four $M_1$ observables. In addition, differences between ${l'_1}^\infty$ calibrations are considered of higher orders. 
Specifically, see Tab.~\ref{table_all_M1_l2_0}, the ratios of the NLO to LO contribution are found to be in the range of $0.05-0.13$. In addition, since $\hat{\mu}_d$ has a vanishing NLO contribution, its deviation from the experiment can be regarded as N$^2$LO, and assuming a natural convergence, we expect the ratio of this contribution to the LO contribution to be $({\text{N}^2\text{LO}}/{\text{LO}})\approx ({\text{NLO}}/{\text{LO}})^2$, or ${\text{NLO}}/{\text{LO}}\approx 0.1$. Finally, the different calibration methods of ${l'_1}^\infty$ (see Tab.~\ref{table_all_M1_l2_0}) lead to a variation in the predictions for the different ${l'_1}^\infty$-dependent observables. This variation represents the contribution from higher orders. Thus, the ratio of the variation to the NLO contribution should be of the order of the expansion parameter. Using Tab.~\ref{table_all_M1_l2_0}, this leads to $({\text{N}^2\text{LO}}/{\text{NLO}})\approx 0.04-0.1$. 

If one assumes that the expansion parameter $\delta$ is common to all $M_1$ observables, one can use the results to assess the value of $\delta$. In order to do this, let us take the log average of all the aforementioned estimates of the expansion parameter: $\log\left| a^{\text{N}^k\text{LO}}_{M_1}/a^{\text{N}^{k-1}\text{LO}}_{M_1}\right|=\log\delta+\log R$. The numbers $R$ are positive natural numbers and are not biased; they, therefore, should be distributed about ``1'', 
and a sum over the logarithms of the different ``R''s should vanish. The log average of many such estimates should converge to $\log \delta$. The fact that this is a finite-size sample means that there remains a measure of uncertainty in determining $\delta$, represented as a distribution. We find that at a 95\% degree of belief, the expansion parameter is within the range of $0.05<\delta<0.13$. The above suggests that the expansion converges faster, by more than a factor of 3 than the na\"ive \pilesseft estimate.

\begin{figure}[h!]
 \centering
 \includegraphics[width=1\linewidth]{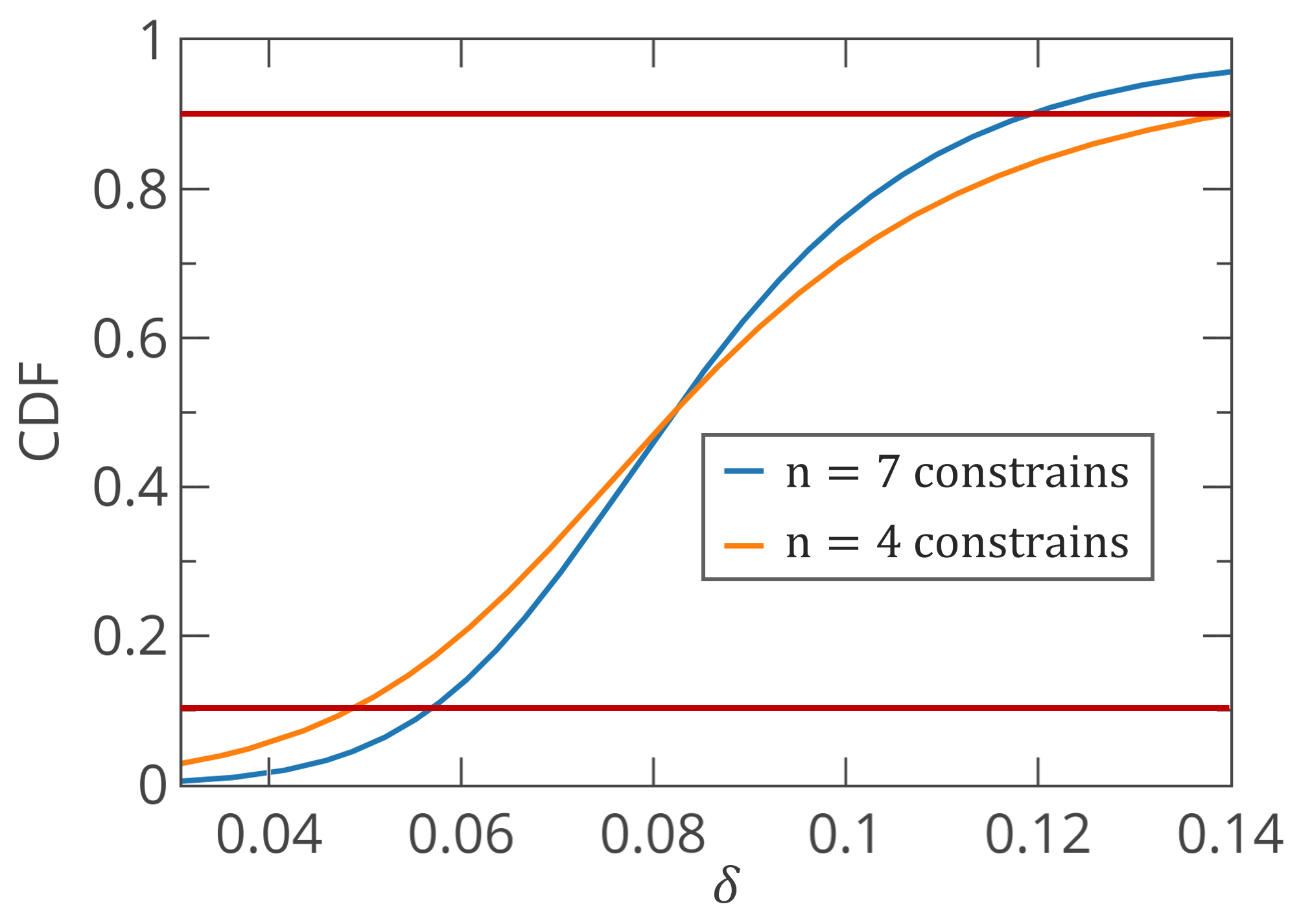}\\
 \caption{\footnotesize{
  Cumulative Density Functions (CDFs) of $\delta$, the expansion parameter. The blue curve represents a calculation that takes into account the constraints of the NLO contributions of $\langle\hat{\mu}_{^3\text{H}}\rangle$, $\langle\hat{\mu}_{^3\text{He}}\rangle$, $Y_{np}$, the N$^2$LO contribution of $\langle\hat{\mu}_d\rangle$, and the variation of ${l’_1}^\infty$. The orange curve takes into account only the first four constraints. The red lines limit the $10\%-90\%$ probability range.}} 
 \label{fig_delta_CDF2}
\end{figure}

In order to check the sensitivity of the expansion parameter to the number of observables, we calculate the Cumulative Density Functions (CDFs) of $\delta$, the expansion parameter, with all the $n=7$ constraints: the NLO contributions of $\langle\hat{\mu}_{^3\text{H}}\rangle$, $\langle\hat{\mu}_{^3\text{He}}\rangle$, ${Y}_{np}$, the N$^2$LO contribution of $\langle\hat{\mu}_d\rangle$, and the variation of ${l’_1}^\infty$ stems from the three electromagnetic observables. Also, we calculate the CDF of $\delta$ only with the $n=4 $ first constraints stemming from the order of the calculation and not from the LEC variation. As shown in Fig.~\ref{fig_delta_CDF2}, with a 70\% degree of belief, the effect of the change is rather small (a change of about 20\% in the estimated truncation error). At higher degrees of belief, especially above 90\%, the truncation error significantly depends upon the number of constraints, as can be expected due to the small number of such constraints.

The truncation error of an expansion, given a prior which represents the naturalness of the expansion, follows a posterior that was calculated in Refs.~\cite{Griesshammer:2015ahu, PhysRevC.92.024005, Cacciari2011}. In the current case, since the expansion parameter is unknown,\textit{ i.e.,} follows the prior distribution in Fig.~\ref{fig_delta_CDF2}, one should fold these two distributions to find the posterior distribution of these two distributions. The formalism is further explained in Appendix~A. 

\begin{figure}[h!]
 \centering
 \includegraphics[width=1\linewidth]{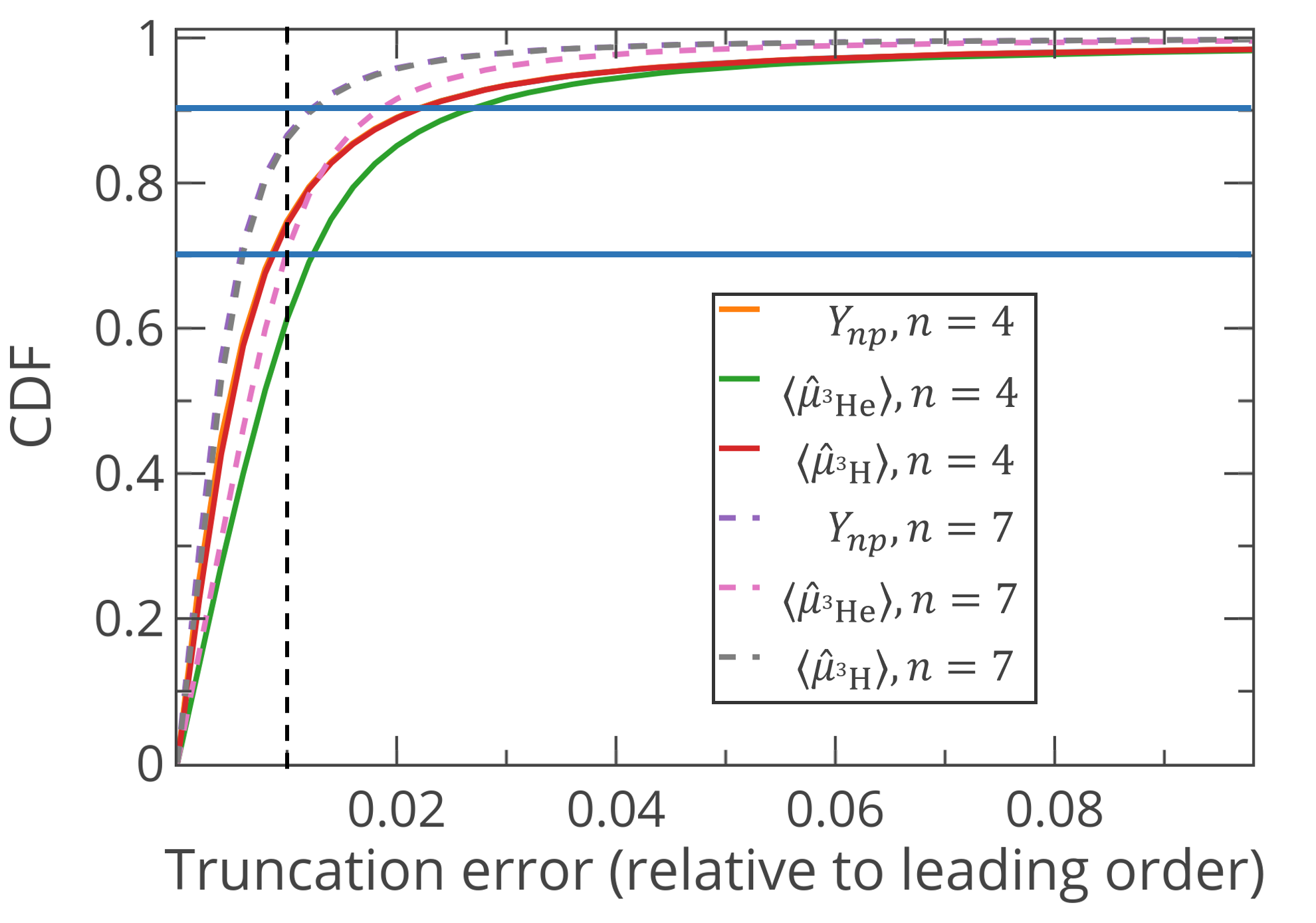}\\
 \vspace{1 cm}
 \caption{\footnotesize{
  Cumulative Density Functions (CDFs) for the different observables, as calculated using \cref{eq_delta}. Horizontal lines are the 70\% and 90\% degrees of belief. We show CDFs relevant to expansion parameter priors with $n=4$ (solid lines) and $n=7$ (dashed lines) constraints, as explained in Fig.~\ref{fig_delta_CDF2}. }} 
 \label{fig_delta_CDF}
\end{figure}

Fig.~\ref{fig_delta_CDF} shows that at about a 70\% degree of belief, the theoretical uncertainty of the calculated $M_1$ observables is about 1\%~\footnote{The uncertainty varies with the observable. 1\% is the maximal value, to be on the conservative side.}.

\subsection{Final results compared to experiment}

\begin{figure}[h!]
\includegraphics[width=\linewidth]{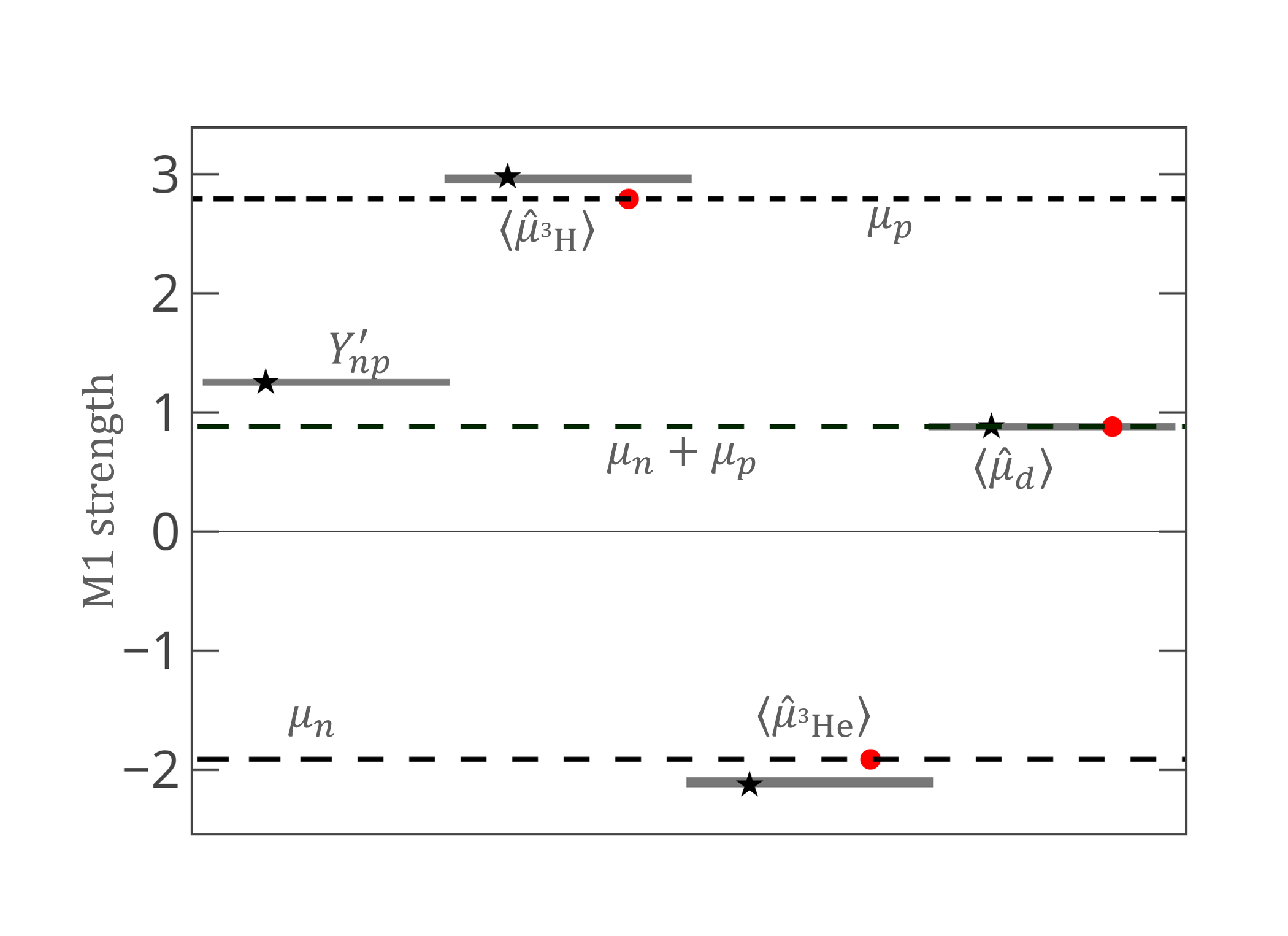}\\
\caption{\footnotesize{The strength of $A=2, \, 3$ $M_1$ observables with a 70\% degree of belief. The bands correspond to the theoretical Z-parameterization theoretical uncertainty from the calculations shown Fig.~\ref{fig_A_2_3}. The stars are the experimental values while the dots are the shell model prediction.}} 
\label{fig_A_2_3}
\end{figure}

Using $Z$-parameterization, our final predictions for the electromagnetic interactions are, to 70\% degree of belief:
\begin {table}[H]
\begin{center}
 \begin{tabular}{c c c c c}
 \centering
 &&This work [{NM}]& Experiment $[{\rm NM}]$&\\ 
 $Y'_{np}$ &=& $1.253\pm0.006$ & $1.2532 \pm 0.0019$&\cite{np_data} \\
 $\langle\hat{\mu}_{^3\text{H}}\rangle$ &=& $2.99\pm0.015$ & $2.97896...$ &\cite{3He_3H_data} \\
 $\langle\hat{\mu}_{^3\text{He}}\rangle$ &=& $ -2.11\pm0.02 $ & $-2.12750...$&~\cite{3He_3H_data}\\
 $\langle\hat{\mu}_d\rangle$& =&$0.88\pm0.01$&0.857~&~\cite{mu_d_data}~,
 \end{tabular}
\end{center}
\end{table} where the uncertainty for each $M_1$ observable is estimated from our calculation (see Fig.~\ref{fig_delta_CDF2}). These results are visually shown in Fig.~\ref{fig_A_2_3}, where for each observable, the bands correspond to the theoretical uncertainty in the Z-parameterization calculation, as calculated in the previous sub-section. 

\section{Discussion} \label{discussion}

The results presented in the previous section indicate that \pilesseft has a predictive power. The prescription we presented, i.e., an NLO calculation in the Z-parameterization, augmented with an uncertainty estimate, reaches percent level accuracy and precision, showing an excellent agreement with the experimental results within a quantitatively estimated 1\% theoretical uncertainty. 

This is an amazing result, considering the small number of parameters in the model, as well as their nature. All parameters in the nuclear model, but the three-body forces, have a concrete experimental origin, i.e., scattering lengths and effective ranges, with no additional fit or correction needed. \cblack{Since three-body forces are calibrated from the 3-body nuclei binding energy, the four-nucleon-one-magnetic-photon low energy constants are the only fit parameters}. \cblack Since there are four unknown $M_1$ low energy observables, this is a verification and validation of the calculation method.

However, we find empirically two facts that do not coincide with the na\"ive \pilesseft expansion. The first fact is that the isoscalar coupling of two nucleons to the magnetic photon is found to be consistent with a vanishing value. In terms of EFT, one can interpret this fact as an indication that the isoscalar two-body magnetic coupling is of a higher order than NLO. 

Let us speculate regarding a physical origin for such suppression. One of the heralds of QCD at low energies is the spontaneous chiral symmetry breaking, resulting in the appearance of an isovector Nambu-Goldstone boson, i.e., the pion. A low-energy EFT that takes this into account is Chiral EFT ($\chi$EFT), in which viable degrees of freedom include not only the nucleon but also the pion. Two-body coupling with magnetic currents in $\chi$EFT, see, e.g., \cite{Park:1994sr,Kolling:2011mt,Pastore:2012rp,Bacca:2014tla,Pastore:2008ui}, include also an explicit pion exchange. As the characteristic momentum transfer decreases, exchanged pions become static and can in principle, be integrated out to a contact interaction. Indeed, the isovector nature of the pion entails that the leading two-body currents are of isovector nature (at NLO), while isoscalar is further suppressed to N$^3$LO. 

\pilesseft is constructed usually from symmetry considerations, but, in principle, it is also a Renormalization Group flow of QCD, and its EFTs, towards low-energies. Though an explicit flow of $\chi$EFT into \pilesseft has not been explicitly demonstrated, we conjecture that the suppression of the isoscalar two-body coupling is a result of this suppression in $\chi$EFT, hence originating in the isovector character of the pion. To our knowledge, this is the first time a signature of $\chi$EFT is found to affect the power counting in a \pilesseft calculation. Here, we are able to pin-point this due to the statistical type of analysis that shows that the standard deviation in ${l'_2}^\infty$ is of the order of its value. 

Recently, a large $N_c$ analysis \cite{Richardson:2020iqi} has also shown that the isoscalar two-body magnetic coupling is suppressed, by about a factor of 20, compared to the isovector two-body magnetic coupling. This alternative explanation additionally shows that this suppression is natural at the QCD scale but highlights the aforementioned questions regarding the special renormalization group flow towards low-energy for magnetic operators. 

The other surprising fact is the unnaturally small expansion parameter $\delta=5\%-10\%\ll\frac{\gamma_t}{m_\pi}\approx\frac{1}{3}$. The fact that the NLO contribution is so small is traced to the shell structure of the magnetic moments of the A=2,\,3 bound nuclei. A natural expansion parameter would have created $\approx 30\%$ deviation from the shell model prediction. We remind that the \pilesseft prediction of the quantum structure of the nuclei is inconsistent with the shell model for these nuclei. Also, we point that for the $^3$H $\beta$ decay (which is the weak analogous for $\langle \mu_{^3\text{H}}\rangle$ and $\langle \mu_{^3\text{He}}\rangle$) $\delta$ was found to be $~3\%$ \cite{De-Leon:2016wyu}, with $~6\%$ deviation from the shell model prediction, which is consistent with our estimations for $\delta$. 

As the expansion parameter is reminiscent of the breakdown scale, a small expansion parameter may hint that this low-energy EFT is actually representing different expansion physics. 

For example, K\"onig et al.~\cite{konig5, Konig:2016utl} have shown that many features of the structure of nuclei emerge from a strictly perturbative expansion about the unitary limit, which is accompanied by a different expansion parameter: $\aleph_0/Q\sim10\%-15\%$.
Using the unitary limit, i.e., $1/\left|a_t\right|\ll Q\ll\Lambda_{\rm b}$, they were able to calculate physical observables such as $^3$H and $^3$He binding energy up to NLO, which are very similar to those obtained using a physical value of $a_t$. 
The similar expansion parameter, which we empirically find in the current work, might hint toward a common physical origin.

A different explanation can be related to an expansion about a leading Wigner-SU(4) symmetry, in which the nucleon is an SU(4) symmetry object, the isovector, and spin SU(2) sub-groups. This has been suggested by Vanasse and Phillips \cite{Vanasse2017}, who examine $^3$H and $^3$He point charge radius in this limit. They indeed found that $\delta$ is smaller than the na\"{i}ve \pilesseft estimate. Our result can be related to this suggestion, in particular since one notices that using the Z-parameterization is close to the SU(4) limit, as $\rho_t\approx\rho_s$. 

Vanasse and Phillips further point out that the emergence of SU(4) symmetry in \pilesseft calculations is surprising, as this symmetry is usually emergent at the QCD breakdown scale, i.e., $\Lambda_{\text{QCD}}\approx 1 \,\text{GeV}$, and thus it is not trivial that it will survive the Renormalization Group flow below the pion mass scale. Here we note that such a property would also explain the survival through Renormalization Group Flow of the suppression of the isoscalar two-body coupling. Such a property of the flow might be related to the specific operator, which in the $M_1$ case can be traced back to SU(4) generators. 

Nuclear EFT expansions, and their power counting, are still a matter of debate in the literature, so it is of interest to further study these suggestions. 

\section{Summary}\label{summary}
In this paper, we present a detailed study of $A=2,\ 3$ $M_1$ observables using \pilesseft up to NLO, analytically built and numerically verified to be RG invariant by order-by-order consistent and controlled perturbative expansion, to describe the structure and dynamics of the reactions, making use of the low characteristic momentum of the reactions and the involved nuclei.

We check two different NLO arrangements, {\it, i.e.,} ER-parameterization, which uses at NLO, the value of the $^3S_1$ effective range, and Z-parameterization, which fixes the Deuteron pole position exactly at NLO. In both cases, the next-to-leading order contribution amounts to less than 10\% correction, which is smaller than the na\"ive expansion parameter of \pilesseft.

The four observables are used to fix, in six different ways, ${l' _1}^\infty$ and ${l' _2}^\infty$, two unknown NLO LECs representing two-nucleon electromagnetic isovector and isoscalar nuclear currents, respectively.

This process leads to an unnaturally small NLO contribution, which in the case of the magnetic moments, means that the deviation from shell model predictions is small. Thus, \pilesseft recovers the shell structure only due to the unexpectedly small expansion parameter. 
Moreover, in both parameterizations, we find that the
correction to their matrix element originating from the
two-body isoscalar low-energy constant, ${l'_2}^\infty$,
is consistent with zero, again in contrast to
the na\"ive dimensional analysis of pionless EFT. Together with the small expansion parameter, these are two surprising deviations from the na\"ive \pilesseft expectation, which hint towards properties of the renormalization group flow of QCD into \pilesseft low-energy regime. The small expansion parameter might be a result of SU(4) symmetry, which in principle exists in EFTs with a breakdown scale of $\sim 1\,\text{GeV}$, where the shell-model structure is a LO effect. A vanishing isoscalar two-body contribution at NLO is a property of chiral symmetry due to the isovector character of the pion. The fact that these properties survive the renormalization group flow to low energies is probably an observable specific property that is of interest in the context of understanding power counting and nuclear EFTs expansions, and so the $M_1$ observables can serve as a starting point for studies of power counting systematics in nuclear EFTs.

We further judge the validity of the NLO parameterizations through the fluctuations in the values of the predicted observables, using the resulting spectrum of LEC values fixed by different choices of observables.
The Z-parameterization is found to have a natural convergence pattern and very stable results. The ER-parameterization, however, is found to have large fluctuations in the predicted
results. We, therefore, focused on the Z-parameterization.

We demonstrate that by using the $Z$-parameterization, the values of the short-range strengths are consistent in the $A=2$ and $A=3$ systems, showing no need for a three-body current. We develop a Bayesian approach (\cblack {which will describe in more detail in future work\cite{D._fake}}) \cblack estimate the theoretical uncertainty due to the truncation of the EFT expansion. This is found to be about 1\% for the calculated observables. The results reproduce to high precision the experimental values of the $M_1$ observables within a 70\% degree of belief band.

We found that \pilesseft predictions for the electromagnetic observables have unprecedented precision and accuracy, comparable with $\chi$EFT calculations~\cite{Pastore:2008ui}, which have about a factor of 3 more parameters and lack the robust uncertainty estimate and consistency we give here. These, as summarized in Fig.~\ref{fig_A_2_3}, verify and validate the way we applied \pilesseft at NLO.

Hence, this calculation opens a new path to model-independent calculations of low-energy electromagnetic and weak reactions, including reactions taking place in the interior of stars.

\begin{acknowledgments}
We thank Daniel Phillips and Harald Grie{\ss}hammer for their comments on the work summarized in this paper. Sebastian K$\ddot{\text{o}}$nig, Jared Vanasse, and Johannes Kirscher are thanked, as
well as the rest of the participants of the GSI-funded EMMI RRTF
workshop ER15-02: Systematic Treatment of the Coulomb Interaction in
Few-Body Systems, for valuable discussions that have contributed significantly to completing this work. The research was supported by ARCHES and by the ISRAEL SCIENCE FOUNDATION
(grant No. 1446/16). 
\end{acknowledgments}

\section*{Appendix A - A Bayesian approach to estimate the convergence rate} %
\setcounter{subsection}{0} 
\setcounter{equation}{0} 
\setcounter{section}{0} 
\setcounter{figure}{0}
\renewcommand{\thesection}{A}
\renewcommand{\theequation}{A-\arabic{equation}}
\renewcommand{\thesubsection}{A.\Roman{subsection}}
\renewcommand{\thefigure}{A.\arabic{figure}}
\label{ap_Bayesian approach}
In the following, we expand upon the approach we used to evaluate the truncation uncertainty and the size of the expansion parameter of a set of observables whose expansion is $\langle{M_1}\rangle=\langle{M_1}\rangle_{\text LO} \cdot \left ( 1+c^{\text NLO}_{M_1} \delta + c^{\text N2LO}_{M_1} \delta^2+{\mathcal{O}} (\delta^3)\right)$.

\subsection{The Bayesian probability distribution of the expansion parameter}
We use information theory arguments to understand the expected behaviour of the expansion convergence rate. The ratio of the k$^{th}$ and l$^{th}$ expansion terms should be proportional to $\delta^{k-l}$ ($\delta$ is the expansion parameter), \textit{i.e.}, 
\begin{equation}
r_{k-l}^{M_1}\delta^{k-l}\equiv\left|\frac{c^{{\text {N}}^k\text{LO}}_{M_1}}{c^{{\text {N}}^l\text{LO}}_{M_1}}\right|.
\end{equation}
\pilesseft formalism suggests that $r_{k-l}^{M_1}$ should be a natural number. We interpret this as a statement regarding the nature of the distribution of these numbers. In layman's terms, one would be surprised if these numbers deviate much from 1. In other words, these coefficients have some natural range of change $\frac{1}{\alpha} < r_{k-l}^{M_1} < \alpha$, where $\alpha$ is a measure of naturalness. One can expect $\alpha$ a factor of 2-3, while bigger variations are acceptable as long as they are rare. From a Bayesian point of view, $r_{k-l}^{M_1}$ are independent and identically distributed random variables (i.i.d) with an average of about $1$ and their logarithm has a (unknown) standard deviation of $\log \alpha$. 

Information theory now states that the probability density function (pdf) $f(r)$ should maximize the entropy $S[f]=-\int dr f(r)\log f(r)$ subject to the constraints $\overline{\log r} =0$ and $\overline{ (\log r-\overline{\log r})^2}=\log \alpha$.
Thus, the log-average of $r_{k-l}^{M_1}\delta^{k-l}$ should be the expansion parameter $(k-l)\log{\delta}$. These lead to a pdf $f(r)$ that is a log-normal distribution.

One can now use a sample of the size $n$ to estimate the expansion parameter, i.e., $\overline{\log{\delta}}=\frac{1}{n}\sum_{i=1}^n \frac{1}{k-l} \log\left( r_{k-l}^{M_1(i)}\delta^{k-l}\right) $. Then, by using Bayes theorem, the resulting distribution for the expansion parameter is Student's {\it t}-distribution with n-1 degrees of freedom $\frac{\log\delta-\overline{\log{\delta}}}{\overline{\sigma^2}/\sqrt{n}}\sim T (\overline{\sigma^2}, n-1)$. 

\subsection{The Bayesian probability distribution of the truncation error of an expansion, given a prior for the expansion parameter}

A good estimate for the truncation error is the maximal coefficient in an expansion of order $k$, multiplied by $\delta^{k+1}$. In Refs.~\cite{Griesshammer:2015ahu, PhysRevC.92.024005, Cacciari2011}, a Bayesian probability distribution is calculated for the truncation error under the assumption of a natural expansion, albeit in the case where the expansion parameter $\delta$ is known. In what follows, we combine their idea with the probability distribution for the expansion parameter found above, to find the Bayesian probability distribution that the NLO result will deviate by $\Delta$ from the true value. Then, 
\begin{align}\label{eq_delta}
\nonumber
&pr\left (\Delta \big | \left\{a^{\text NLO}_{M^k_1}\right\}_{k=1}^n\right)=\\
&\int d\delta pr\left (\Delta \big | \left\{c^{\text NLO}_{M^k_1}\right\}_{k=1}^n, \delta\right)\cdot pr\left (\delta \big | \left\{a^{\text NLO}_{M^k_1}\right\}_{k=1}^n\right).
\end{align}
$pr (\Delta \big | \left\{c^{\text NLO}_{M^k_1}\right\}_{k=1}^n, \delta)$ is calculated in~Ref.~\cite{Griesshammer:2015ahu}, and at NLO, is roughly constant for $|\Delta|\le R_\xi (\delta)$, and decays as $1/|\Delta|^3$ for $|\Delta|\ge R_\xi (\delta)$, where $R_\xi (\delta)=max\left (1, \left\{c^{\text NLO}_{M^k_1}\right\}_{n=1}\right)\delta^2$. The exact functional form depends on the prior assumption for $\left\{c^{\text NLO}_{M^k_1}\right\}_{k=1}^n$, but a common behavior of all the checked priors is that at a degree of belief of $\frac{k}{k+1}$ (translating to $\approx 67\%$ for NLO calculations) the resulting truncation error is less than $R_\xi (\delta) \delta^{k+1}$. 

As the pdf for $\delta$, we take the Student's {\it t}-distribution found in the previous subsection.

\section*{Appendix B - Calculating the $M_1$ of $A=3$ bound states }\label{ap_magnetic_matrix_element} %

\setcounter{equation}{0} 
\setcounter{section}{0} 
\setcounter{subsection}{0} 
\setcounter{figure}{0}
\renewcommand{\thesection}{B}
\renewcommand{\theequation}{B-\arabic{equation}}
\renewcommand{\thesubsection}{B.\Roman{subsection}}
\renewcommand{\thefigure}{B.\arabic{figure}}
In this appendix, we present the general method for calculating an
		an $A=3$ matrix element in \pilesseft and its application for calculating $M_1$ observables of an $A=3$ system. 
		
		The $A=3$ magnetic moments are defined as matrix elements between $A=3$ bound sate wave functions of $\psi^{^3\text{H}}$, $\psi^{^3\text{He}}$, using the general formalism introduced in Ref.~\cite{Big_paper}.
		\begin{equation}\label{eq_general_operator}
		\langle \mathcal{O}\rangle=a^J\bra{{S, S'_{z}, I, I'_{z}, E',q} }\mathcal{O}^{J}\mathcal{O}^{I}\mathcal{O}^q\ket{S, S_{z}, I, I_{z}, E},
		\end{equation}
		
		Where $a^J$ originates from the reduction of the multipole operator (See~\cite{Walecka:1995mi}) and :
		\begin{itemize}
		\item $\langle \frac{1}{2}, S_{z}, J, m_z| \frac{1}{2}, S'_{z} \rangle \neq 0$
		\item $I'_{z}= \begin{cases}
		-I_{z}& \mathcal{O}^I=\tau^{\pm}\\
		I_{z}& \mathcal{O}^I=\tau^0
		\end{cases}$
		\end{itemize}
		where $\tau^{\pm}$ is isospin ladder operators. $\psi_{i,(j)}$ is the initial (final) three-nucleon wave function as defined in Ref.~\cite{Big_paper}.
		\subsubsection{$A=3$, matrix element one-body operator}
		In Ref.~\cite{Big_paper}, we showed that at LO, the three-nucleon normalization can be written as:
		\begin{equation}
		1= \sum\limits_{\mu,\nu}
		\bra{\psi^{i}_\mu}
		\mathcal{O}_{\mu\nu}^{\text{norm}}(E_i)\ket{\psi_\nu^{i}}~,
		\end{equation}
		where $\mathcal{O}_{\mu\nu}^{\text{norm}}(E_i)$ is the normalization operator such that: 
		\begin{multline}\label{eq:1b:Onorm}
		\mathcal{O}_{\mu\nu}^{\text{norm}}(E_i)=\\
		\frac{\partial}{\partial_E}\left [\hat{I}_{\mu\nu}(E,p,p')-My_\mu y_\nu a^i_{\mu\nu}K_{\mu\nu}^i(p',p,E)\right]\bigg|_{E=E_i}~,
		\end{multline}
		where:
		\begin{equation}
		K_{\mu\nu}^i=\begin{cases}
		K_0(p',p,E)&i=^3\text{H}\\
		K_0(p',p,E)+K_{\mu\nu}^C(p',p,E)&i=^3\text{He}
		\end{cases}
		\end{equation}
		and
		\begin{equation}\label{eq_a_mu_nu_3H}
		a_{\mu\nu}=\begin{array}{c|cc}
		\mbox{\backslashbox{$\mu$\kern-1em}{\kern-1em$\nu$}}&t&s \\ \hline
		t&-1&3\\
		s&3&-1\\
		\end{array}~,
		\end{equation}
		\begin{equation}\label{eq_a_mu_nu_3He}
		a'_{\mu\nu}=\begin{array}{c|ccc}
		\mbox{\backslashbox{$\mu$\kern-1em}{\kern-1em$\nu$}}&t&np&pp \\ \hline
		t&-1&3&3\\
		np&1&1&-1\\
		pp&2&-2&0
		\end{array}~,
		\end{equation}
		\cblack
		
		are a result of the different projection operators (see Ref.~\cite{Parity-violating} for example) and we have defined the operation:
		\begin{equation}\label{eq_otimes}
		A (..., p)\otimes B (p, ...)=\int A (.., p)B (p, ...)\frac{p^2}{2\pi^2}dp~.
		\end{equation}
		\begin{eqnarray}
		\hat{I}_{\mu\nu}(p,p',E)&=&\frac{2\pi^2}{{ p}^2}\delta\left(p-p'\right)D_{\mu}(E,p)^{-1}\delta_{\mu,\nu}\\
		K_0 (p,p',E)&=&\frac{1}{2pp'}Q_0\left (\frac{p^2+p'^2-ME}{pp'}\right)~,
		\end{eqnarray}
		where $\delta_{\mu,\nu}$ is the Kronecker delta and:
		\begin{gather}
		\label{Q}
		Q_0 (\text{a})=\frac{1}{2} \int^1_{-1}\frac{1}{x+a}dx~.
		\end{gather}
		$K_{\mu\nu}^C(p'',p',E)$ is the $\mu,\nu$ index of the one-photon exchange matrix, $K^C (p'', p', E)$ (see Ref.~\cite{Big_paper}), $\mu,\nu =t,s$ are the different triton channels, $\mu =t,s,pp$ are the different $^3$He channels, $y_{\mu,\nu}$ are the nucleon-dibaryon coupling constants for the different channels, $a_{\mu\nu}$ ($a'_{\mu\nu}$) are a result of the $n-d$ ($p-d$) doublet-channel projection (\cite{Parity-violating}) and $D_{\mu}(E,p)$ is the dibaryon propagator (\cite{Big_paper,Triton,3bosons}).
		
		A general one-body operator, can be written as a generalization of a three-nucleon normalization operator for the case of both energy and momentum transfer, between initial (i) and final
		(j) $A=3$ bound-state wave functions ($\psi_{i,j})$. The general operator $\mathcal{O}_{j,i}$ is factorizes into the
		\cblack following parts:
		\begin{equation}
		\mathcal{O}_{j,i}=\mathcal{O}^{J}\mathcal{O}^{T}\mathcal{O}_{j,i}(q_0,q), 
		\end{equation}
		where $\mathcal{O}^{J}$, the spin part of the operator whose total
		spin is $J$, and $\mathcal{O}^{T}$, the isospin part of the operator,
		that depend on the initial and final quantum numbers. The spatial part
		of the operator, $\mathcal{O}_{j,i}(q_0,q)$, is a function of the
		three-nucleon wave function's binding energies, ($E_i$,$E_j$) and the energy and
		momentum transfer ($q_0,q$, respectively).
		In the case of a triton $\beta$-decay, the spin and isospin one-body operators are combinations of Pauli matrices, so their reduced matrix element ($\langle \| F\|\rangle$, $\langle \| GT\|\rangle$) can be easily calculated as a function of the three-nucleon quantum total spin and isospin numbers. In Ref.~\cite{Big_paper} we showed that the reduced matrix element of such an operator can be written as:
		\begin{multline}\label{eq_general_operator_reduced}
		\langle\| \mathcal{O}_{j,i}^{\text{1-B}}(q_0,q)\|\rangle=\\
		\left\langle\frac{1}{2}\left\|\mathcal{O}^J\right\|\frac{1}{2}\right\rangle\left\langle\frac{1}{2},I'_z\left|\mathcal{O}^T\right|I_z,\frac{1}{2}\right\rangle\\ \times
		\sum\limits_{\mu,\nu}
		\bra{\psi^j_\mu}y_\mu y_\nu\Bigl\{d'^{ij}_{\mu\nu} \hat{\mathcal{I}}(q_0,q)\\+
		a'^{ij}_{\mu\nu}\left [\hat{\mathcal{K}}(q_0,q)+{\hat{\mathcal{K}}^C_{\mu\nu}}(q_0,q)\right]\Bigr\} \ket{\psi^i_\nu} 
		~,
		\end{multline}
		such that for $i=j$:
		\begin{eqnarray}
		d'^{ii}_{\mu\nu}&=&\delta_{\mu,\nu}\\
		a'^{ii}_{\mu\nu}&=&\begin{cases}
		a_{\mu\nu}&i=j=^3\text{H}\\
		a'_{\mu\nu}&i=j=^3\text{He}\\
		\end{cases}.
		\end{eqnarray}
		The spatial parts of operator are are denoted by
		$\hat{\mathcal{I}}(E, q_0,q)$, $\hat{\mathcal{K}}(q_0,q)$ and ${\hat{\mathcal{K}}^C_{\mu\nu}}(E, q_0,q)$. The full analytical
		expressions for $\hat{\mathcal{I}}(E, q_0,q)$ and $\hat{\mathcal{K}}(E, q_0,q)$
		are given in Ref.~\cite{Big_paper} while ${\hat{\mathcal{K}}^C_{\mu\nu}}(q_0,q)$
		are the diagrams that contain a one-photon interaction in addition to
		the energy and momentum transfer. A derivation of an analytical
		expression for these diagrams is too complex, so they were calculated
		numerically only. $a'^{ij}_{\mu\nu}$ and $d'^{ij}_{\mu\nu}$ are
		a result of the $N-d$ doublet-channel projection coupled to
		$\mathcal{O}^J\mathcal{O}^T$ (for more details, see Ref~\cite{Big_paper}). 
		\subsubsection{Two-body matrix element}
		In contrast to the normalization operator given in \cref{eq:1b:Onorm},
		which contains only one-body interactions, a typical \pilesseft
		electroweak interaction contains also the following two-body
		interactions up to NLO:
		\begin{equation}
		t^\dagger t,\,s^\dagger s,\,(s^\dagger t+h.c)~,
		\end{equation}
		under the assumption of energy and momentum conservation. 
		The diagrammatic form of the different two-body interactions is given in Ref.~\cite{Big_paper}.
		
		\subsection{magnetic $A=3$ matrix element}
		Based on Ref.~\cite{Big_paper}, the three-nucleon magnetic moment matrix element that contains \textbf{one-body} interactions, $\hat{\mu}^{(\text{1-B})}$, can be written as:
		\begin{multline}\label{eq_one_body}
		\langle \hat{\mu}^{(\text{1-B})}\rangle=\dfrac{\left\langle\frac{1}{2}\left\|\boldsymbol{\sigma}\right\|\frac{1}{2}\right\rangle}{\sqrt{6}}\\ \times
		\sum\limits_{\mu,\nu}\int\frac{d^3p}{(2\pi)^3}\int\frac{d^3p'}{(2\pi)^3}
		{\psi^j_\mu(p')} y_\mu y_\nu\left(\kappa_0+\kappa_1\tau_3\right)
		\\ \times\left[d'^{ij}_{\mu\nu} \hat{\mathcal{I}}(0,0)
		+ a'^{ij}_{\mu\nu}\hat{\mathcal{K}}(0,0)\right] {\psi^i_\nu(p)} ~,
		\end{multline}
		where the full analytical
		expressions for the direct and exchange spatial operators, $\hat{\mathcal{I}}(E, q_0,q)$ and $\hat{\mathcal{K}}(E, q_0,q)$,
		are given in Ref.~\cite{Big_paper}.
		
		For $^3$H:
		\begin{equation}\label{eq_d_mu_3H}
		d'^{ii}_{\mu\nu}=\begin{array}{c|ccc}
		\mbox{\backslashbox{$\nu$\kern-1em}{\kern-1em$\mu$}}&t&np&nn \\
		\hline
		t&\frac{2}{3}\mu_n+\frac{1}{3}\mu_p&\mu_n-\mu_p&0\\
		np&\frac{\mu_n-\mu_p}{3}&\mu_p&0\\
		nn&0&0&\mu_n
		\end{array}~,
		\end{equation}
		and \begin{equation}\label{eq_a_3H}
		\mbox{ $a'^{ii}_{\mu\nu}$}=\begin{array}{c|ccc}
		\mbox{\backslashbox{$\mu$\kern-1em}{\kern-1em$\nu$}}&t&np&nn \\ \hline
		t&-\left(\frac{5}{3}\mu_p-\frac{2}{3}\mu_n\right)&\mu_p+2\mu_n&3\mu_p\\
		np&\frac{2}{3}\mu_n+\frac{1}{3}\mu_p&2\mu_n-\mu_p&-\mu_p\\
		nn&2\mu_p&-2\mu_p&0
		\end{array}~,
		\end{equation}
		and for $^3$He:
		\begin{equation}\label{eq_d_3He}
		d'^{ii}_{\mu\nu}=\begin{array}{c|ccc}
		\mbox{\backslashbox{$\nu$\kern-1em}{\kern-1em$\mu$}}&t&np&nn \\
		\hline
		t&\frac{2}{3}\mu_p+\frac{1}{3}\mu_n&\mu_p-\mu_n&0\\
		np&\frac{\mu_p-\mu_n}{3}&\mu_n&0\\
		nn&0&0&\mu_p
		\end{array}
		\end{equation}
		and \begin{equation}\label{eq_a_mu_3He}
		\mbox{ $a'^{ii}_{\mu\nu}$}=\begin{array}{c|ccc}
		\mbox{\backslashbox{$\mu$\kern-1em}{\kern-1em$\nu$}}&t&np&nn \\ \hline
		t&-\left(\frac{5}{3}\mu_n-\frac{2}{3}\mu_p\right)&\mu_n+2\mu_p&3\mu_n\\
		np&\frac{2}{3}\mu_p+\frac{1}{3}\mu_n&2\mu_p-\mu_n&-\mu_n\\
		nn&2\mu_n&-2\mu_n&0
		\end{array}~,
		\end{equation}
		
		Note that for $^3$H and $^3$He, \cref{eq_one_body} is defined such that for the special case of equal neutron and proton magnetic moments ($\mu_n=\mu_p=\kappa_0$):
		\begin{multline}\label{eq_one_body_1}
		\langle \hat{\mu}^{(\text{1-B})}\rangle=\kappa_0
		\sum\limits_{\mu,\nu}\int\frac{d^3p}{(2\pi)^3}\int\frac{d^3p'}{(2\pi)^3}
		{\psi^j_\mu(p')} y_\mu y_\nu\\
		\times\left[d'^{ii}_{\mu\nu} \hat{\mathcal{I}}(0,0)
		+a'^{ii}_{\mu\nu}\hat{\mathcal{K}}(0,0)\right] {\psi^i_\nu(p)} =\kappa_0~,
		\end{multline}
		since:
		\begin{multline}\sum\limits_{\mu,\nu}\int\frac{d^3p}{(2\pi)^3}\int\frac{d^3p'}{(2\pi)^3}
		{\psi^j_\mu(p')} y_\mu y_\nu\\
		\times\left[d'^{ii}_{\mu\nu} \hat{\mathcal{I}}(0,0)
		+a'^{ii}_{\mu\nu}\hat{\mathcal{K}}(0,0)\right]{\psi^i_\nu(p)} =1~,
		\end{multline}
		is just the normalization condition of the bound state (see Ref.~\cite{Big_paper}). \cblack

		\begin{figure}[h!]
			\centering
			\includegraphics[width=\linewidth]{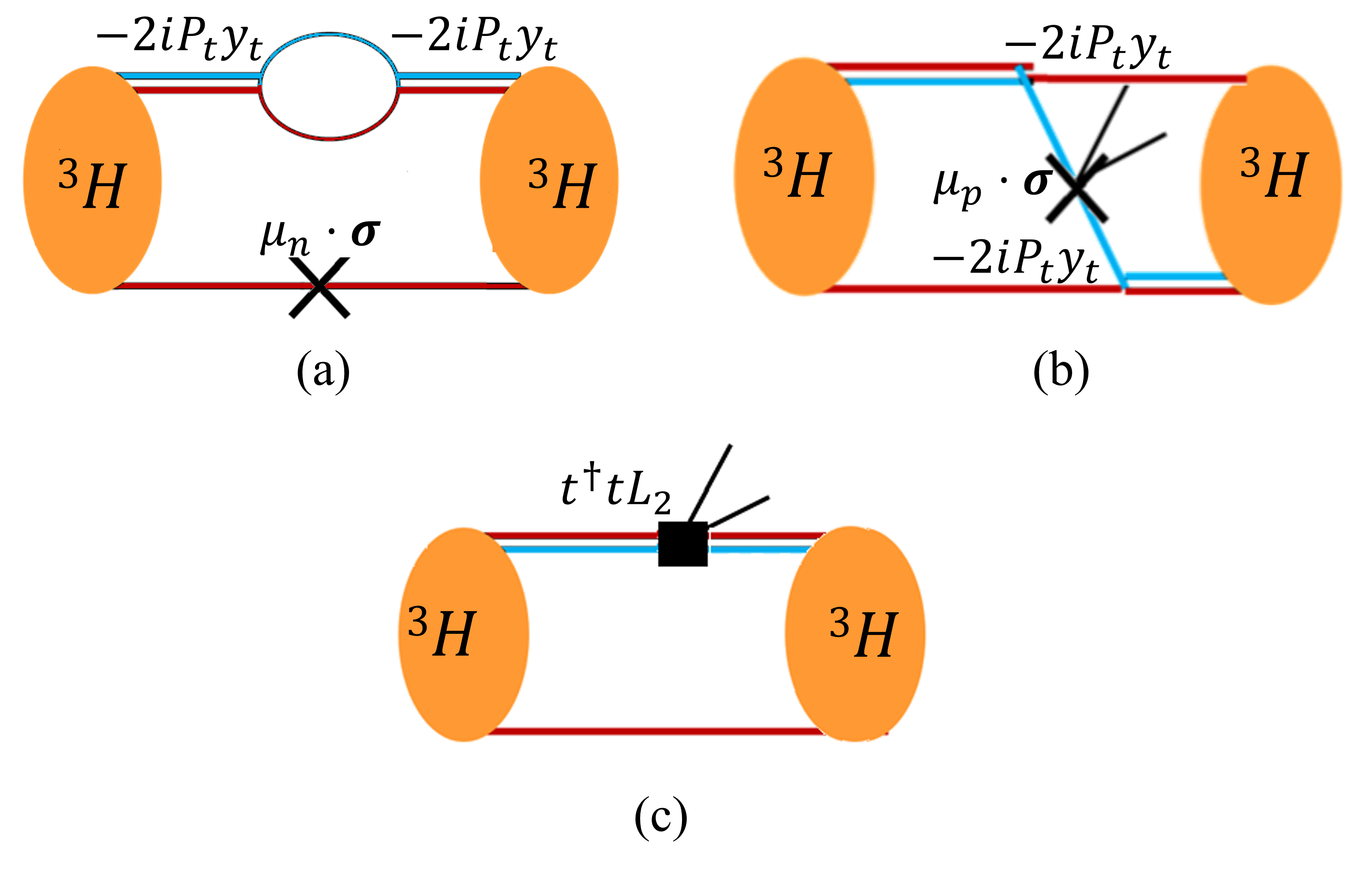}\\
			\caption{\footnotesize{ Three of the diagrams contributing to the triton magnetic moment. 
					The one-body diagrams are coupled to $\mu_n\cdot \boldsymbol{\sigma}$ (a) and $\mu_p\cdot \boldsymbol{\sigma}$ (b), diagram (c) is coupled to the two-body LEC $L_2$. The double lines are the propagators of the dibaryon field $D_t$ (solid). The red lines represent the neutron propagator, while the blue lines represent the proton propagator.}}\label{fig_magnetic_intercation}
	\vspace{1 cm}	
	\end{figure}
	
			For example, Fig.~\ref{fig_magnetic_intercation} presents a one-body $A=3$ matrix element coupled to $i\frac{e}{2M}\left(\kappa_0+\kappa_1\tau_3\right)\boldsymbol{\sigma},
		$
		$\mu_n$ in the case of neutron and 
		$\mu_p$ in the case of proton.
		
		At LO, Fig.~\ref{fig_magnetic_intercation} (a) is given by: 
		\begin{multline}
		\mu_ny_t^2\int\dfrac{d^3 p}{(2\pi)^3}\int\dfrac{d^3p'}{(2\pi)^3}\dfrac{2\pi^2}{p^2}\\ \times \Biggl[\psi^{^3\text{H}}_t(p)
		\dfrac{M^2}{4 \pi \sqrt{3 p^2-4ME_{^3\text{H}}}}\delta(p-p')
		\psi^{E_{^3\text{H}}}_t(p')\Biggr]~.
		\end{multline}
		Fig.~\ref{fig_magnetic_intercation} (b) is given by: 
		\begin{align}
		\nonumber
		&\mu_py_t^2\int\dfrac{d^3 p}{(2\pi)^3}\int\dfrac{d^3p'}{(2\pi)^3}\psi^{^3\text{H}}_t(p)\\ \times
		&\frac{M^2}{p^2 \left(p^2-2 ME_{^3\text{H}}\right)+\left(p^2-ME_{^3\text{H}}\right)^2+p'^4}
		\psi^{E_{^3\text{H}}}_t(p')~.
		\end{align}

		The three-nucleon magnetic moment matrix element that contains \textbf{two-body} interactions, $\hat{\mu}^{(\text{2-B})}$, can be written as the sum of all two-body interactions (see Ref.~\cite{Big_paper}): 
		\begin{multline}
		\langle\hat{\mu}^{(\text{2-B})} \rangle =\\\int\frac{d^3p}{(2\pi)^3}\left\{\frac{2}{3}L'_2\psi^2_t(p)+L_1'a^{(2)}_{ts}\left[\psi_t(p)\times\psi_{np}(p)+h.c\right]\right\},
		\end{multline}
		where
		\begin{equation}
		a^{(2)}_{ts}=
		\begin{cases}-\frac{2}{3} &^3\text{H}\\
		1 &^3\text{He}.
		\end{cases}
		\end{equation}
		
		Fig.~\ref{fig_magnetic_intercation} (c) is given by: 
		\begin{align}
		&L_2\int\dfrac{d^3 p}{(2\pi)^3}\left[\psi^{^3\text{H}}_t(p)\right]^2.
		\end{align}
	\bibliography{references}

\begin{thebibliography}{10}

\bibitem{Beane:2014ora}
S.~R. Beane, E.~Chang, S.~Cohen, W.~Detmold, H.~W. Lin, K.~Orginos, A.~Parreno,
  M.~J. Savage, and B.~C. Tiburzi.
\newblock {Magnetic moments of light nuclei from lattice quantum
  chromodynamics}.
\newblock {\em Phys. Rev. Lett.}, 113(25):252001, 2014.

\bibitem{Beane:2013br}
S.~R. Beane et~al.
\newblock {Nucleon-Nucleon Scattering Parameters in the Limit of SU(3) Flavor
  Symmetry}.
\newblock {\em Phys. Rev.}, C88(2):024003, 2013.

\bibitem{Bethe:1949yr}
H.~A. Bethe.
\newblock {Theory of the Effective Range in Nuclear Scattering}.
\newblock {\em Phys. Rev.}, 76:38--50, 1949.

\bibitem{3bosons}
Paulo~F. Bedaque, H.~W. Hammer, and U.~van Kolck.
\newblock {The Three boson system with short range interactions}.
\newblock {\em Nucl. Phys.}, A646:444--466, 1999.

\bibitem{Triton}
P.~F. Bedaque, H.~W. Hammer, and U.~van Kolck.
\newblock {Effective theory of the triton}.
\newblock {\em Nucl. Phys.}, A676:357--370, 2000.

\bibitem{KSW1998_a}
D.~B. Kaplan, M.~J. Savage, and M.~B. Wise.
\newblock {Two nucleon systems from effective field theory}.
\newblock {\em Nucl. Phys.}, B534:329--355, 1998.

\bibitem{KSW1998_b}
D.~B. Kaplan, M.~J. Savage, and M.~B. Wise.
\newblock {A New expansion for nucleon-nucleon interactions}.
\newblock {\em Phys. Lett.}, B424:390--396, 1998.

\bibitem{Griesshammer_pionless}
H.~W. Griesshammer.
\newblock {Naive dimensional analysis for three-body forces without pions}.
\newblock {\em Nucl. Phys.}, A760:110--138, 2005.

\bibitem{few_platter}
L.~Platter.
\newblock {Few-Body Systems and the Pionless Effective Field Theory}.
\newblock {\em PoS}, CD09:104, 2009.

\bibitem{PhysRevLett.115.132001}
S.~R. Beane, E.~Chang, W.~Detmold, K.~Orginos, Assumpta Parre\~no, M.~J.
  Savage, and B.~C. Tiburzi.
\newblock Ab initio.
\newblock {\em Phys. Rev. Lett.}, 115:132001, Sep 2015.

\bibitem{faddeev}
L.D. Faddeev and S.P. Merkuriev.
\newblock {\em Quantum Scattering Theory for Several Particle Systems}.
\newblock Springer, 1993.

\bibitem{Kong1}
X.~Kong and F.~Ravndal.
\newblock {Proton proton fusion in leading order of effective field theory}.
\newblock {\em Nucl. Phys.}, A656:421--429, 1999.

\bibitem{konig2}
J.~Vanasse, D.~A. Egolf, John Kerin, S.~K{\"o}nig, and R.~P. Springer.
\newblock {$^{3}$He and $pd$ Scattering to Next-to-Leading Order in Pionless
  Effective Field Theory}.
\newblock {\em Phys. Rev.}, C89(6):064003, 2014.

\bibitem{K_nig_2017}
Sebastian König.
\newblock Second-order perturbation theory for
  {\textdollar}$\lbrace$$\rbrace${\^{}}$\lbrace$3$\rbrace${\textbackslash}mathrm$\lbrace$he$\rbrace${\textdollar}
  andpdscattering in pionless {EFT}.
\newblock {\em Journal of Physics G: Nuclear and Particle Physics},
  44(6):064007, may 2017.

\bibitem{Big_paper}
H.~De-Leon, L.~Platter, and D.~Gazit.
\newblock Calculation of $a = 3$ bound state matrix element in pionless
  effective field theory.
\newblock 2019.

\bibitem{3He_3H_data}
J.~E. Purcell, J.~H. Kelley, E.~Kwan, C.~G. Sheu, and H.~R. Weller.
\newblock {Energy levels of light nuclei {A} = 3}.
\newblock {\em Nucl. Phys.}, A848:1--74, 2010.

\bibitem{mu_d_data}
P.~J. Mohr and B.~N. Taylor.
\newblock {CODATA recommended values of the fundamental physical constants:
  1998}.
\newblock {\em Rev. Mod. Phys.}, 72:351--495, 2000.

\bibitem{np_data}
AE~Cox, SAR Wynchank, and CH~Collie.
\newblock The proton-thermal neutron capture cross section.
\newblock {\em Nuclear Physics}, 74(3):497--507, 1965.

\bibitem{Bacca:2014tla}
S.~Bacca and S.~Pastore.
\newblock {Electromagnetic reactions on light nuclei}.
\newblock {\em J. Phys.}, G41(12):123002, 2014.

\bibitem{PhysRevC.99.034005}
R.~Schiavilla, A.~Baroni, S.~Pastore, M.~Piarulli, L.~Girlanda, A.~Kievsky,
  A.~Lovato, L.~E. Marcucci, Steven~C. Pieper, M.~Viviani, and R.~B. Wiringa.
\newblock Local chiral interactions and magnetic structure of few-nucleon
  systems.
\newblock {\em Phys. Rev. C}, 99:034005, Mar 2019.

\bibitem{Chen:1999bg}
J.-W. Chen and M.~J. Savage.
\newblock {$np\rightarrow d \gamma$ for big bang nucleosynthesis}.
\newblock {\em Phys. Rev.}, C60:065205, 1999.

\bibitem{Rupak:1999rk}
G.~Rupak.
\newblock {Precision calculation of n p ---> d gamma cross-section for big bang
  nucleosynthesis}.
\newblock {\em Nucl. Phys.}, A678:405--423, 2000.

\bibitem{KSW_c}
D.~B. Kaplan, M.~J. Savage, and M.~B. Wise.
\newblock {A Perturbative calculation of the electromagnetic form-factors of
  the deuteron}.
\newblock {\em Phys. Rev.}, C59:617--629, 1999.

\bibitem{Chen_N_N}
J.-W. Chen, G.~Rupak, and M.~J. Savage.
\newblock {Nucleon-nucleon effective field theory without pions}.
\newblock {\em Nucl. Phys.}, A653:386--412, 1999.

\bibitem{ando_deturon}
S.-I. Ando and Chang~Ho Hyun.
\newblock {Effective field theory on the deuteron with dibaryon field}.
\newblock {\em Phys. Rev.}, C72:014008, 2005.

\bibitem{Kirscher:2017fqc}
Johannes Kirscher, Ehoud Pazy, Jonathan Drachman, and Nir Barnea.
\newblock {Electromagnetic characteristics of $A \leq 3$ physical and lattice
  nuclei}.
\newblock {\em Phys. Rev.}, C96(2):024001, 2017.

\bibitem{Vanasse:2017kgh}
J.~Vanasse.
\newblock {Charge and Magnetic Properties of Three-Nucleon Systems in Pionless
  Effective Field Theory}.
\newblock {\em Phys. Rev.}, C98(3):034003, 2018.

\bibitem{rearrange}
S.~R. Beane and M.~J. Savage.
\newblock {Rearranging pionless effective field theory}.
\newblock {\em Nucl. Phys.}, A694:511--524, 2001.

\bibitem{Bedaque:1999vb}
P.o~F. Bedaque and H.~W. Griesshammer.
\newblock {Quartet S wave neutron deuteron scattering in effective field
  theory}.
\newblock {\em Nucl. Phys.}, A671:357--379, 2000.

\bibitem{Phillips_N_N}
D.~R. Phillips, G.~Rupak, and M.~J. Savage.
\newblock {Improving the convergence of N N effective field theory}.
\newblock {\em Phys. Lett.}, B473:209--218, 2000.

\bibitem{Griesshammer_3body}
H.~W. Griesshammer.
\newblock {Improved convergence in the three-nucleon system at very low
  energies}.
\newblock {\em Nucl. Phys.}, A744:192--226, 2004.

\bibitem{Kong2}
X.~Kong and F.~Ravndal.
\newblock {Proton proton fusion in effective field theory}.
\newblock {\em Phys. Rev.}, C64:044002, 2001.

\bibitem{Phillips:1999am}
D.~R. Phillips and T.~D. Cohen.
\newblock {Deuteron electromagnetic properties and the viability of effective
  field theory methods in the two nucleon system}.
\newblock {\em Nucl. Phys.}, A668:45--82, 2000.

\bibitem{Vanasse}
J.~Vanasse.
\newblock {Fully Perturbative Calculation of $nd$ Scattering to
  Next-to-next-to-leading-order}.
\newblock {\em Phys. Rev.}, C88(4):044001, 2013.

\bibitem{Vanasse:2015fph}
Jared Vanasse.
\newblock {Triton charge radius to next-to-next-to-leading order in pionless
  effective field theory}.
\newblock {\em Phys. Rev.}, C95(2):024002, 2017.

\bibitem{Rho:1999bv}
M.~Rho.
\newblock {Effective field theory for nuclei, dense matter and the Cheshire
  Cat}.
\newblock {\em AIP Conf. Proc.}, 494:391--399, 1999.

\bibitem{33}
J.~J. de~Swart, C.~P.~F. Terheggen, and V.~G.~J. Stoks.
\newblock {The Low-energy n p scattering parameters and the deuteron}.
\newblock 1995.

\bibitem{Preston_1975}
M.~A. Preston and R.~K Bhaduri.
\newblock {\em Structure of the nucleus, by M. A. Preston and R. K. Bhaduri}.
\newblock Addison-Wesley Pub. Co., Advanced Book Program, 1975.

\bibitem{deSwart:1995ui}
J.~J. de~Swart, C.~P.~F. Terheggen, and V.~G.~J. Stoks.
\newblock {The Low-energy n p scattering parameters and the deuteron}.
\newblock 1995.

\bibitem{34}
J.~R. Bergervoet, P.~C. van Campen, W.~A. van~der Sanden, and Johan~J.
  de~Swart.
\newblock {Phase shift analysis of 0-30 MeV pp scattering data}.
\newblock {\em Phys. Rev.}, C38:15--50, 1988.

\bibitem{Kaplan:1998xi}
D.~B. Kaplan, M.~J. Savage, R.~P. Springer, and M.~B. Wise.
\newblock {An Effective field theory calculation of the parity violating
  asymmetry in (polarized) $n + p \rightarrow d + \gamma$}.
\newblock {\em Phys. Lett.}, B449:1--5, 1999.

\bibitem{ando_magntic_BBN}
S.~Ando, R.~H. Cyburt, S.~W. Hong, and C.~H. Hyun.
\newblock {Radiative neutron capture on a proton at BBN energies}.
\newblock {\em Phys. Rev.}, C74:025809, 2006.

\bibitem{Chen:1999tn}
J.-W. Chen, G.~Rupak, and M.~J. Savage.
\newblock {Nucleon-nucleon effective field theory without pions}.
\newblock {\em Nucl. Phys.}, A653:386--412, 1999.

\bibitem{Note1}
The purpose of this work is to examine the consistency of pionless EFT for the
  transition from $A=2$ to $A=3$ and vice versa. Hence, to eliminate any
  artificial effects, we choose to use the same value $\left (\mu = \Lambda
  \rightarrow \infty \right )$ for both the two-body dimensional regularization
  and the three-body cutoff regularization, since we expect that $\protect
  \qopname \relax m{lim}_{\mu = \Lambda \rightarrow \infty }\protect \frac {d
  M_1(\mu )}{d\mu }=0$ for both A=2 and A=3 observables.

\bibitem{Note2}
In this work, similar to Refs.~\cite {Vanasse:2017kgh,PhysRevLett.115.132001},
  we use $q=2Mv_{\protect \text {lab}}=2\cdot M 2200 \protect \text
  {M/s}=0.0069\protect \tmspace +\thinmuskip {.1667em} \protect \text {MeV}$.
  This value is higher than the value used in Refs.~\cite
  {Chen:1999tn,ando_deturon},$q=0.068\protect \tmspace +\thinmuskip {.1667em}
  \protect \text {MeV}.$ We found that $l_1' (q=0.0069\protect \tmspace
  +\thinmuskip {.1667em} \protect \text {MeV})$ is higher than $l_2'
  (q=0.0068\protect \tmspace +\thinmuskip {.1667em} \protect \text {MeV})$ by
  20\%(10\%) for the Z-(ER-) parameterization.

\bibitem{3H_3He_magnetic}
R.~G. Sachs and Julian Schwinger.
\newblock {The Magnetic Moments of H-3 and He-3}.
\newblock {\em Phys. Rev.}, 70:41--43, 1946.

\bibitem{Note3}
Here, since we are focusing on the ERE order-by-order expansion of the matrix
  element, we took the full Coulomb descriptions of $^3$He. It should be noted
  that in another work that focused on weak interaction \cite {de2019tritium},
  we examined the effects of Coulomb interaction on the weak matrix element and
  showed thing at it is significantly lower in comparison to that originates
  from the ERE.

\bibitem{2016PhLB..755..253K}
J.~{Kirscher} and D.~{Gazit}.
\newblock {The Coulomb interaction in Helium-3: Interplay of strong short-range
  and weak long-range potentials}.
\newblock {\em Physics Letters B}.

\bibitem{Note4}
Cutoff dependence for low cutoff is expected since the cutoff functions
  inflicted on matrix element and wave functions cannot be made fully
  consistent.

\bibitem{PhysRevC.92.024005}
R.~J. Furnstahl, N.~Klco, D.~R. Phillips, and S.~Wesolowski.
\newblock Quantifying truncation errors in effective field theory.
\newblock {\em Phys. Rev. C}, 92:024005, Aug 2015.

\bibitem{Griesshammer:2015ahu}
H.~W. Griesshammer, J.~A. McGovern, and D.~R. Phillips.
\newblock {Nucleon Polarisabilities at and Beyond Physical Pion Masses}.
\newblock {\em Eur. Phys. J.}, A52(5):139, 2016.

\bibitem{Cacciari2011}
M.~Cacciari and N.~Houdeau.
\newblock Meaningful characterisation of perturbative theoretical
  uncertainties.
\newblock {\em Journal of High Energy Physics}, 2011(9):39, Sep 2011.

\bibitem{Note5}
The uncertainty varies with the observable. 1\% is the maximal value, to be on
  the conservative side.

\bibitem{Park:1994sr}
T.-S. Park, D.-P. Min, and M.~Rho.
\newblock {Radiative neutron - proton capture in effective chiral Lagrangians}.
\newblock {\em Phys. Rev. Lett.}, 74:4153--4156, 1995.

\bibitem{Kolling:2011mt}
S.~Kolling, E.~Epelbaum, H.~Krebs, and U.~G. Meissner.
\newblock {Two-nucleon electromagnetic current in chiral effective field
  theory: One-pion exchange and short-range contributions}.
\newblock {\em Phys. Rev.}, C84:054008, 2011.

\bibitem{Pastore:2012rp}
S.~Pastore, Steven~C. Pieper, R.~Schiavilla, and R.~B. Wiringa.
\newblock {Quantum Monte Carlo calculations of electromagnetic moments and
  transitions in {A} $\leq$ 9 nuclei with meson-exchange currents derived from
  chiral effective field theory}.
\newblock {\em Phys. Rev.}, C87(3):035503, 2013.

\bibitem{Pastore:2008ui}
S.~Pastore, R.~Schiavilla, and J.~L. Goity.
\newblock {Electromagnetic two-body currents of one- and two-pion range}.
\newblock {\em Phys. Rev.}, C78:064002, 2008.

\bibitem{Richardson:2020iqi}
Thomas~R. Richardson and Matthias~R. Schindler.
\newblock {Large-$N_c$ analysis of magnetic and axial two-nucleon currents in
  pionless effective field theory}.
\newblock 2 2020.

\bibitem{De-Leon:2016wyu}
Hilla De-Leon, Lucas Platter, and Doron Gazit.
\newblock {Tritium $\beta$-decay in pionless effective field theory}.
\newblock {\em Phys. Rev. C}, 100(5):055502, 2019.

\bibitem{konig5}
S.~K{\"o}nig, H.~W. Grießhammer, H.~W. Hammer, and U.~van Kolck.
\newblock {Effective theory of $^3$H and $^3$He}.
\newblock {\em J. Phys.}, G43(5):055106, 2016.

\bibitem{Konig:2016utl}
Sebastian König, Harald~W. Grießhammer, H.~W. Hammer, and U.~van Kolck.
\newblock {Nuclear Physics Around the Unitarity Limit}.
\newblock {\em Phys. Rev. Lett.}, 118(20):202501, 2017.

\bibitem{Vanasse2017}
J.~Vanasse and D.~R. Phillips.
\newblock Three-nucleon bound states and the wigner-su(4) limit.
\newblock {\em Few-Body Systems}, 58(2):26, Jan 2017.

\bibitem{D._fake}
D.~Gazit and H.~De-Leon.
\newblock Truncation error estimation and order-order by analysis for pionless
  eft observables.
\newblock {\em In preparation}, 2020.

\bibitem{Walecka:1995mi}
J.~D. Walecka.
\newblock {Theoretical nuclear and subnuclear physics}.
\newblock {\em Oxford Stud. Nucl. Phys.}, 16:1--610, 1995.

\bibitem{Parity-violating}
H.~W. Griesshammer, M.~R. Schindler, and R.~P. Springer.
\newblock {Parity-violating neutron spin rotation in hydrogen and deuterium}.
\newblock {\em Eur. Phys. J.}, A48:7, 2012.

\bibitem{de2019tritium}
Hilla De-Leon, Lucas Platter, and Doron Gazit.
\newblock Tritium $\beta$ decay in pionless effective field theory.
\newblock {\em Physical Review C}, 100(5):055502, 2019.

\end{thebibliography}
\bibliographystyle{unsrt}
\end{document}